\documentclass{aa}  
\usepackage[varg]{txfonts}
\usepackage{threeparttable}
\usepackage{graphicx}
\usepackage{xspace}
\usepackage{amsmath}
\usepackage{placeins}
\usepackage{gensymb}
\usepackage{color}
\usepackage{float}
\usepackage{todonotes}
\usepackage{xcolor}
\usepackage{nicematrix}
\usepackage{ulem}
\usepackage{multicol}
\usepackage{multirow}

\renewcommand{\vec}[1]{\mbox{\boldmath$#1$}}
\newcommand{\music}{\texttt{MUSIC}\xspace}
\newcommand{\Msun}{M_\odot}
\newcommand{\Lsun}{L_\odot}
\newcommand{\tconv}{\tau_{\mathsf{conv}}}
\newcommand{\sigmavp}{\sigma_{\mathsf{vp,in}}}

\newcommand{\rmax}{r_\mathsf{max}}
\newcommand{\ro}{r_\mathsf{o}}
\newcommand{\DPS}{\displaystyle}

\definecolor{orange}{rgb}{.9,.3,0}

\begin{document}
\title{Convective shells in the interior of Cepheid variable stars: overshooting models based on hydrodynamic simulations}
\author{M. Stuck\inst{\ref{inst1}} \and J. Pratt\inst{\ref{inst1},\ref{inst2},\ref{inst3}} \and I. Baraffe\inst{\ref{inst3},\ref{inst4}} \and J.A. Guzik\inst{\ref{inst5}}  \and M.-G. Dethero\inst{\ref{inst2}} \and D.G. Vlaykov\inst{\ref{inst3}} \and T. Goffrey\inst{\ref{inst6}}
\and A. Le Saux \inst{\ref{inst7}} }
\institute{Lawrence Livermore National Laboratory,7000 East Ave, Livermore, CA 94550, USA \label{inst1}
\and Department of Physics and Astronomy, Georgia State University, Atlanta GA 30303, USA \label{inst2} 
\and Department of Physics and Astronomy, University of Exeter, EX4 4QL Exeter, United Kingdom \label{inst3}
\and \'Ecole Normale Sup\'erieure de Lyon, CRAL (UMR CNRS 5574), Universit\'e de Lyon 1, 69007 Lyon, France\label{inst4}
\and Los Alamos National Laboratory, Los Alamos, NM 87545, USA\label{inst5} 
\and Centre for Fusion, Space and Astrophysics, Department of Physics, University of Warwick, Coventry, CV4 7AL, United Kingdom \label{inst6}
\and Université Paris-Saclay, Université Paris Cité, CEA, CNRS, AIM, Gif-sur-Yvette, F-91191, France. \label{inst7}
}

\titlerunning{Convection in Cepheid variable stars}
\authorrunning{M. Stuck et al.}

\abstract
{Because Cepheid variable stars have long been used as a cosmic benchmark for scaling distances in our Galaxy and beyond, the accuracy of stellar evolution models for Cepheids have wide-reaching effects. However, our understanding of the dynamics in the interiors of these physically complex stars is limited.}
{Our goal is to provide a detailed multi-dimensional picture of hydrodynamic convection and convective boundary mixing in the interior of Cepheids.}
{Using the Modules for Experiments in Stellar Astrophysics (MESA), we study the structure of intermediate-mass stars that cross the instability strip. Then, we perform two-dimensional hydrodynamic simulations of six of these stars with the fully compressible Multidimensional Stellar Implicit Code (\music). Our simulations do not model the radial pulsations but focus on the interior structure of this family of stars. We develop and apply a new statistical analysis to examine convection and convective boundary mixing in the interior of these stellar simulations.}
{
Based on a grid of MESA models, we demonstrate that a common structure for intermediate mass Cepheids includes an interior convective shell as well as a thin outer convective envelope. Using the extreme value theory approach to analyse our \music simulation data, we find that overshooting above the convective shell fills the space between these convectively unstable layers.  We develop a new statistical analysis that provides a clearer picture of how overshooting fills this layer, and also allows us to formulate a detailed comparison between overshooting above and below the convective shell.  Our analysis effectively decomposes the overshooting layer into two layers: a weak overshooting layer and a strong overshooting layer.  Statistically, this is accomplished by decomposing the strongly
non-Gaussian probability density function into a mixture of Gamma distributions. 
 Using our mixture model, we show that the ratio of overshooting lengths above and below the convective shell depends directly on the radial extent of the convective shell as well as its depth in the star. We propose a new form for the diffusion coefficient, which addresses the need for overlapping overshooting layers between convective shells. We introduce the idea of ``super-mixing layer'' where overshooting from both the convective shell and the convective envelope results in efficient mixing and could be viewed as merging the two adjacent convective zones.
}
{}
\maketitle



\section{Introduction \label{sec:intro} }

Since the discovery of the fundamental relationship between their pulsation period and their luminosity \citep[the Leavitt Law,][]{leavitt1912periods}, classical Cepheids have had a large influence on  astronomy and astrophysics.  Bright Cepheids are easily detected both inside and outside the galaxy, because the period of their radial pulsations typically ranges from a few days to a few weeks \citep[e.g.,][]{bono2005classical, anderson2014effect}. Measurements and models for these variable stars provide a foundation for studies as diverse as the spiral structure of the Milky Way \citep[e.g.][]{majaess2009characteristics, dambis2015classical}, extra-galactic distances \citep[see reviews ][]{feast1987cepheids,feast1999cepheids}, and estimation of the Hubble constant \citep{ngeow2006hubble, van2007cepheid, tammann2008expansion, freedman2010hubble, freedman2021measurements}. Because of the Period--Luminosity (PL) relationship, the study of Cepheids has been an active research area over several decades \citep{madore1991cepheid, turner2010pl, bono2010insights}, including work on the dependence of their properties on metallicity  \citep{alibert1999period, persson2004new}. 

Despite their importance to astronomers, a theoretical picture of the interior of Cepheids is incomplete \citep[see the recent review by][]{guzik2023challenges}. One clear indication of the limitations of current models is the mass discrepancy problem. A typical disagreement of $10\%$ to $20\%$ is found when comparing the masses of Cepheids obtained with stellar evolution models and the ones calculated based on stellar pulsation measurements \citep{bono2006stellar, keller2008cepheid, pietrzynski2010dynamical, cassisi2011classical}. Dynamical masses of Cepheids in binary systems \citep{evans1997mass, evans2024components, gallenne2018geometrical} are lower than the masses provided by evolution models, and more consistent with pulsation calculations \citep[e.g.][]{stobie1969cepheid, neilson2011cepheid}, highlighting that improvement of current stellar evolution models of Cepheids is a high priority. 

\begin{figure}[t!]
    \centering
    \includegraphics[width=\columnwidth]{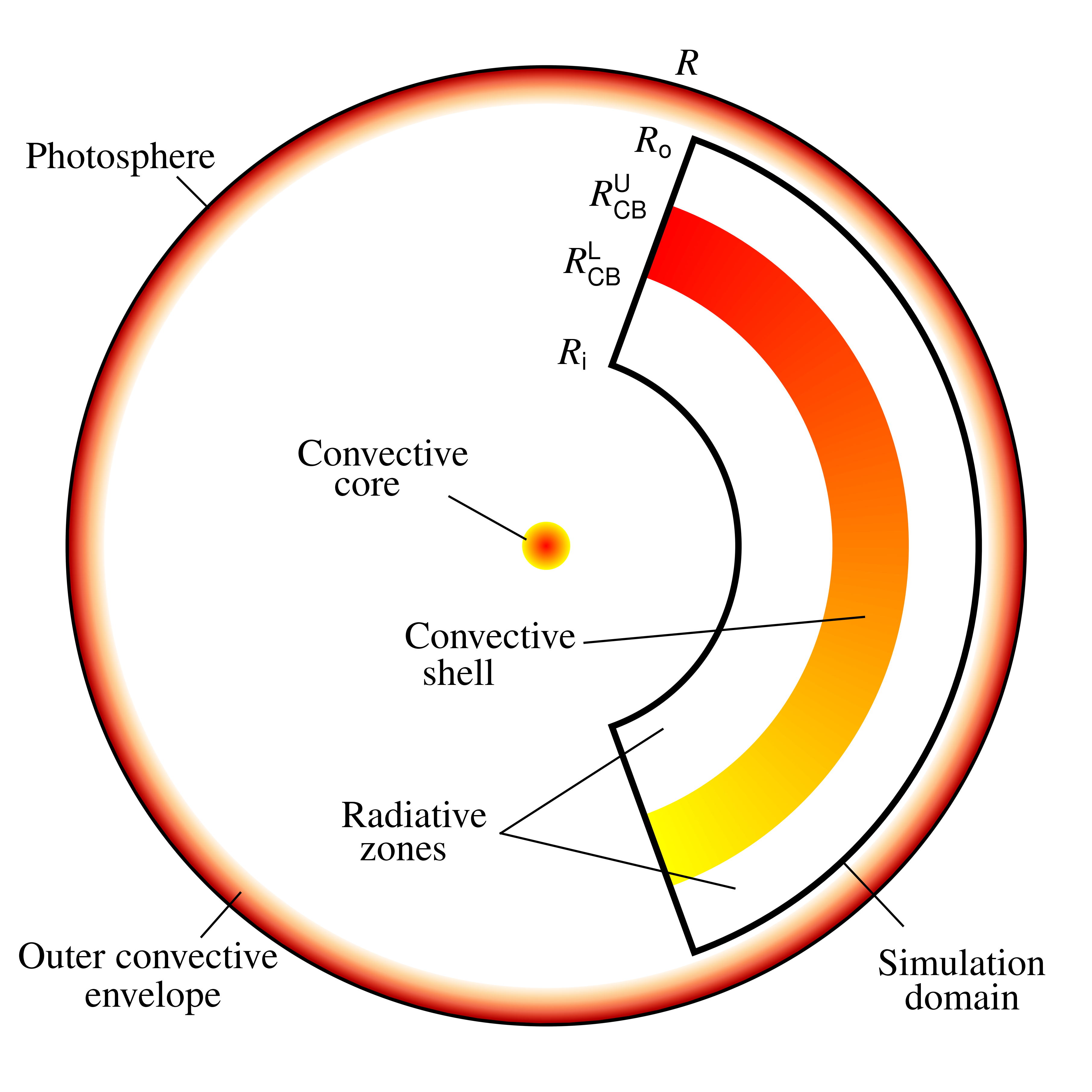}
    \caption{Schematic of the typical structure of the Cepheids we study in this work with \music. Convective regions are colored, including: a tiny convective core, an inner convective shell surrounded by radiative zones, and a thin outer convective envelope. The simulation domain is a spherical shell, and is outlined with black; it encapsulates the interior convection zone along with the surrounding radiative zones, and is truncated just below the outer convective envelope.}
    \label{sketch}
\end{figure}

In the stellar interior, convection underpins the fundamental processes of heat transport, fluid mixing, and shear.  One-dimensional (1D) parameterizations of convection are thus central to stellar evolution models.  The most widely used method to model convection is stellar mixing length theory (MLT) \citep{vitense1953wasserstoffkonvektionszone, bohm1958wasserstoffkonvektionszone,salaris2008stellar,salaris2017chemical}. Because MLT relies on a single parameter, the mixing length, it cannot accurately capture the multi-scale, non-local, multi-dimensional, and intermittent nature of turbulent convection.  MLT is used to model evolutionary tracks for Cepheids before they enter the instability strip, however more sophisticated convection models are generally used to evolve stars while they experience radial pulsation.  This is necessary because radial pulsations can happen on a comparable time-scale to the convection \citep[e.g.,][]{deupree1984two, deupree1986mixing,glasner1995convective}, implying that these two effects can have a two-way interaction. 
Observations have recently verified that such an interaction can occur.  A recent study of R Car \citep{rosales2024new}, a M-type Mira variable star with a pulsation period of 314 days that is evolving along the Asymptotic Giant Branch (AGB) found that convective structures grow when the star expands and decrease when the star contracts.  Because classical Cepheids have a relatively short pulsation period compared to other evolved variable stars, they are likely to be affected more strongly by this interaction.
Time-dependent convection (TDC) models have been developed to model convection during radial pulsations \citep[e.g. ][]{marconi2017recent}. The Radial Stellar Pulsation (RSP) functionality of the Modules for Experiments in Stellar Astrophysics (MESA) \citep{paxton2019modules} models nonlinear pulsations of various pulsating stars by including the TDC model described by \citet{kuhfuss1986model}, \citet{wuchterl1998simple}, and \citet{smolec2008convective}.  The hydrodynamic simulations that we present in this work have the potential to contribute to improvements in both the MLT and TDC models as they are applied to the evolution of Cepheids.

In addition to convection, convective overshooting could play a role in the mass discrepancy problem \citep{chiosi1992instability, bono1999classical}. The effects of convective boundary mixing \citep[e.g.,][]{paxton2010modules,freytag1996hydrodynamical,pratt2017extreme} have been modeled as an add-on to stellar MLT in two different ways. One is with a simple overshooting length, expressed as a fraction of the pressure scale height \citep{schwarzschild1975scale} evaluated at the convective boundary; this overshooting model based on a step function defines a region in the stable regions (RZ, radiative zone) where the evolution model assumes full convective mixing. A more precise approach is based on the formulation of a one-dimensional diffusion coefficient which describes the enhanced mixing in the overshooting layer due to convective motions as a smooth function of the star's internal radius \citep[see e.g.][]{freytag2010role, noels2010overshooting, zhang2013convective,pratt2017extreme}.   Like the step-function model, models based on diffusion coefficients require the input of a length scale, and possibly additional parameters, that must be calibrated.  Multidimensional hydrodynamic simulations of convection can provide insights that improve one-dimensional stellar evolution models, show how those models should be used, and how the relevant parameters can be calibrated \citep[for e.g.,][]{freytag1996hydrodynamical,herwig2000evolution,rogers2006numerical, arnett20193d, kapyla2017extended, pratt2017extreme, baraffe2017lithium, pratt2020comparison, kapyla2023convective}.

Work involving multidimensional hydrodynamic simulations of classical Cepheids is limited \citep[see for e.g.][]{mundprecht2013multidimensional, mundprecht2015multidimensional}; in addition, these works study Cepheid models that include only a thin outer envelope that is convectively unstable.   Many Cepheid models involve an internal convective shell that occurs in the Z-bump region, as well as a convective envelope.  In thin convective shells, plumes can overshoot the convection zone both above and below the shell. Convective shells, which are common in stars on the asymptotic giant branch (AGB), have received less attention than convective envelopes.  Stars with convective shells have been investigated in the context of nuclear burning \citep[e.g.][]{meakin2006active, meakin2007turbulent,mocak2011new,rizzuti2022realistic, rizzuti20233d, rizzuti2024shell,georgy20243d}. \cite{cristini20173d, cristini2019dependence} and \cite{rizzuti2022realistic} performed 3D hydrodynamic "box-in-star" simulations of a massive star and found that convective overshooting is weaker at the bottom of the convective shell than at the top with a factor of approximately $1/5$. However, the depth and thickness of the convective shells in previous hydrodynamic studies have not been clearly documented. Weaker overshooting beneath a convective shell makes intuitive sense because the fluid there is denser than the material above the shell, which could result in shallower penetration of convective plumes. The convective shells in AGB stars are often thicker and located deeper within the star than the Cepheids we examine. Using a custom model in MESA, \cite{li2012k, li2017applications} and \cite{guo2021convective} find that for thin convective shells in A-type stars, the extent of the overshooting layer above the shell is approximately twice the extent of the one below. However the difference that these works find between overshooting depths above and below the convective shell raises several important questions about: (1) whether the overshooting lengths above and below a convective shell can be related to one another, (2) what other stellar properties affect this relationship, (3) how models of convection and convective overshooting can be improved for Cepheids.

A high radial resolution is necessary to accurately resolve overshooting; a long time sequence of data is also important for studying overshooting so that well-resolved statistics of this intermittent process can be built.  For these reasons, we choose to study two-dimensional (2D) simulations in this work. Three-dimensional (3D) simulations are desirable, and will be explored in future work.  2D simulations of stellar convection result in higher velocities than three-dimensional (3D) simulations \citep{muthsam1995numerical, meakin2007turbulent, arnett2011turbulent}; however, the difference between these velocities is dependent on the stellar structure. Despite this general difference, at the convective boundaries radial velocities are similar in 2D and 3D simulations. At the radial resolutions, and for the more limited time sequence of data, that can be produced in 3D, the measured overshooting lengths are also similar to the 2D results \citep{pratt2020comparison,dethero2024shape}.

This work is structured as follows. In Section \ref{sec:sim}, we discuss the realistic global fluid simulations that we have performed with the MUlti-dimensional Stellar Implicit Code (\music) for a selection of these stellar structures. In Section \ref{sec:results}, we examine the shape of the convection in the inner convective shell and its overshooting layers. We propose a new statistical analysis that allows for the direct comparison of overshooting above and below a convective shell.  In Section \ref{secconc}, we discuss the implications of our results.

\begin{table}
\begin{center}
\caption{Parameters of stellar structure models produced with MESA for the hydrodynamic simulations.
 \label{mesasumtable}
 }
\begin{tabular}{lcccccccccccccccccccccccccccc}
                     
    & age & $R$& $L$& $T_{\mathsf{eff}}$& $P$

\\ \hline
ceph5               & $8.25$ & $2.95$ &   $1.73$  &  $5.71$ & $4.74$ 
\\ \hline
ceph6               & $7.83$  & $3.95$ &   $2.40$  &  $5.36$ & $7.64$ 
\\ \hline
ceph7a               & $4.96 $& $5.00$ &  $3.30$    &  $5.16$ & $10.39$ 
\\ \hline
ceph7b               & $5.37$ & $5.54$ & $4.14$    &  $5.19$ & $12.41$ 
\\ \hline
ceph8               & $3.33$ &$ 7.38$ &  $5.15$   &  $4.75$ & $19.00$ 
\\ \hline
ceph9               & $2.95$ & $8.46$ &   $9.00$   & $5.10$ & $22.00$ 
\\ \hline

\end{tabular}
\end{center}
    \begin{tablenotes}
    \item[] \textbf{Notes:}  The star's age in $10^7$ years, the radius of the photosphere of the star in $10^{12}$ cm, the luminosity $L$ of the star at the photosphere in $10^3 L_\odot$, the effective temperature of the star $T_{\mathsf{eff}}$ at the photosphere in $10^3$ Kelvin, and the pulsation period $P$ determined with MESA-RSP in days are included. For each simulation the metallicity $z$ is equal to $0.020$ and the helium mass fraction $y$ is set at $0.29$.
    \end{tablenotes}  
\end{table}

\section{Hydrodynamic simulations \label{sec:sim}}

An interior convective shell is a common feature for Cepheids (see Appendix \ref{sec:mesa} for the detailed analysis of a grid of $48$ MESA models). This structure is illustrated in Fig.~\ref{sketch}. Evolutionary models of stars that may evolve into Cepheids are sensitive to the choice of parameters for convection and overshooting; therefore the presence of an additional convective layer is an important consideration for the improvement of stellar structure and evolution modeling of these stars.  Hydrodynamic simulations of Cepheids, or variable stars more broadly \citep{geroux2011radial, geroux2013radial, geroux2014radial, geroux2015radial, mundprecht2013multidimensional, mundprecht2015multidimensional} have not yet examined models that include an interior convective shell; the present work is a first effort to characterize it. 

\begin{figure}[b!]
\begin{center}
\includegraphics[width=\columnwidth]{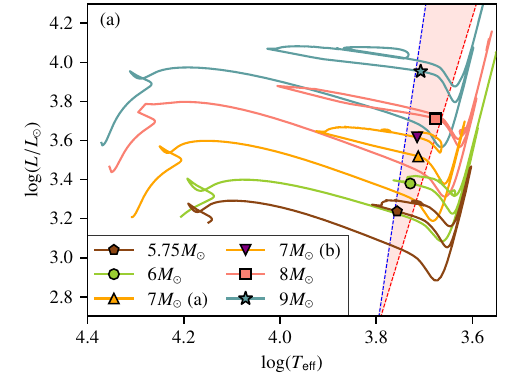}
\includegraphics[width=\columnwidth]{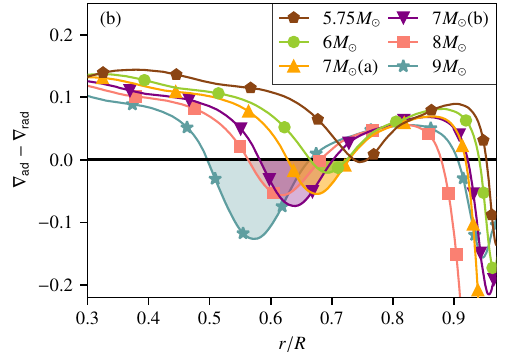}
\caption{(a) HR diagram and (b) radial profile of the Schwarzschild discriminant, for the six stars we simulate with \music. Symbols indicate the Cepheids we are simulating and the pink shaded area indicates the instability strip for the evolutionary tracks. Shaded regions also highlight the internal convection zones around which our hydrodynamic simulations are centered. \label{music_sim_structures}}
\end{center}
\end{figure}

We produce six models of stars of intermediate mass to examine in hydrodynamic simulations; these stars range from $5.75$ to $9\Msun$, and all lie in a blue loop and inside the instability strip.  The MESA inlists for these six stars are based on inlists published by \cite{gilkis2019effects} and \cite{espinoza2022period}. The parameters of the MESA models used in our hydrodynamic simulations are summarized in Table~\ref{mesasumtable}. They have a metallicity of $z=0.020$, which ensures the existence of the deep inner convective shell for all the stellar masses that we consider, and is representative of Cepheids found in the Milky Way. For the $7\Msun$ star, we study two different models, denoted (a) and (b), to allow us a comparison between otherwise similarly structured stars on the second and third crossings of the instability strip. The position of these stars on their evolutionary tracks and the placement of their internal convection zones are provided in Fig.~\ref{music_sim_structures}. Above the internal convective shell, the structure of the Cepheids that we study involves the presence of a thin outer convective envelope. The bottom of this outer convective envelope approximately coincides with the helium-II ionization point, around a temperature of $5\times10^4$K \citep{seaton1993radiative}.  In this outer region, the temperature and density gradients vary more rapidly, requiring significantly higher resolution; including this zone in our simulations with any accuracy would require us to increase the grid sizes by an order of magnitude.  We instead truncate the present simulation volumes at a point just below the outer convective envelope, which allows us to study overshooting above the convective shell without the influence of overshooting below the outer convective envelope.

\begin{table*}[h!]
\begin{center}
\caption{Parameters for compressible hydrodynamic simulations performed with \music.
 \label{simsuma}
 }
\begin{tabular}{lcccccccccccccccccccccccccccc}
                      & $M/M_{\odot}$&  $(R_{\mathsf{i}},R^{\mathsf{L}}_{\mathsf{CB}},R^{\mathsf{U}}_{\mathsf{CB}},R_{\mathsf{o}})/R$  &  $(T^{\mathsf{L}}_{\mathsf{CB}},T^{\mathsf{U}}_{\mathsf{CB}})$($10^5$K)  & $H^{\mathsf{L}}_{\mathsf{p,CB}}/\Delta r$ & $H^{\mathsf{U}}_{\mathsf{p,CB}}/\Delta r$ & $\tconv$($10^6$s) & time ($\tconv$) 
\\ \hline
ceph5              & $5.75$ & $(0.4,0.736,0.759,0.9)$  &   $(1.85,1.65)$ & $126.3$ &   $117.9$    &  $ 4.14 \pm 0.42 $  & $30.13$
\\ \hline
ceph6               & $6$ & $(0.4,0.665,0.725,0.9)$  &   $(2.00, 1.54)$ & $148.1$ &   $130.9$    &  $ 2.00 \pm 0.17 $  & $138.55$
\\ \hline
ceph7a               & $7$ & $(0.45,0.63,0.73,0.85)$  &   $(2.18, 1.47)$  & $202.0$ &  $171.4$    &  $ 2.37 \pm 0.24 $   & $31.06$  
\\ \hline
ceph7b              & $7$ & $(0.3,0.58,0.70,0.85)$  &   $(2.28, 1.44)$  & $205.0$ & $176.0$    &  $ 1.92 \pm 0.21 $  & $25.13$
\\ \hline
ceph8               & $8$ & $(0.2,0.565,0.68,0.8)$  &   $(2.17, 1.43)$  & $242.9$ &  $217.7$    &  $ 2.77 \pm 0.34 $  & $31.02$
\\ \hline
ceph9               & $9$ & $(0.3,0.495,0.665,0.8)$  &   $(2.41,1.31)$  & $207.1$ &   $182.7$    &  $ 2.10 \pm 0.22 $  & $33.48$  
\\ \hline

\end{tabular}
\end{center}
    \begin{tablenotes}
    \item[] \textbf{Notes:}  The simulation name, the inner radius of the spherical shell $R_{\mathsf{i}}$, the radius of the lower convective boundary $R^{\mathsf{L}}_{\mathsf{CB}}$, the radius of the upper convective boundary $R^{\mathsf{U}}_{\mathsf{CB}}$, and the outer radius $R_{\mathsf{o}}$ of the spherical shell are given in units of the total radius of each star, $R$. The temperature is provided at the lower convective boundary $T^{\mathsf{L}}_{\mathsf{CB}}$ and the upper convective boundary $T^{\mathsf{U}}_{\mathsf{CB}}$.  The resolution of the upper (lower) convective boundary $H^{\mathsf{U}}_{\mathsf{p,CB}}/\Delta r$ ($H^{\mathsf{L}}_{\mathsf{p,CB}}/\Delta r$) is provided.  The average global convective turnover time $\tconv$ is provided as well as its standard deviation, and the total time span for each simulation is given in units of the average convective turnover time.
    \end{tablenotes}
\end{table*}

We perform 2D Implicit Large Eddy Simulations (ILES) \citep{grinstein2007implicit,ritos2018performance} of these stars using the \music code.  Our simulations in this work only take convection into account; the possibility of studying additional physical effects such as rotation, a tachocline, chemical mixing, and magnetic fields are omitted from the current study. Of particular relevance is the assumption that each star that we simulate has homogeneous chemical composition. This assumption is consistent with the MESA stellar structures used for these simulations. Because the helium-II ionization point is not included in the simulation volume, radial pulsations should not occur in the present simulations, and we confirm that they do not.  The complex interaction between a star's radial pulsations and convection will be studied in future work.

The \music code solves the inviscid compressible hydrodynamic equations for density $\rho$, momentum $\rho \vec{u}$, and internal energy $\rho e$:
\begin{eqnarray} \label{densityeq}
\frac{\partial}{\partial t} \rho &=& -\nabla \cdot (\rho \vec{u})~,
\\ \label{momeq}
\frac{\partial}{\partial t} \rho \vec{u} &=& -\nabla \cdot (\rho \vec{u} \vec{u}) - \nabla p + \rho \vec{g} ~,
\\ \label{ieneq}
\frac{\partial}{\partial t} \rho e &=& -\nabla \cdot (\rho e\vec{u}) -p \nabla \cdot \vec{u} + \nabla \cdot (\chi \nabla T) ~,
\end{eqnarray}
using a finite volume method, a MUSCL method \citep{thornber2008improved} of interpolation, and a van Leer flux limiter \citep[as described by][]{van1974towards,roe1986characteristic,leveque2006computational}.  Time integration in the \music code is fully implicit and uses a Jacobian free Newton-Krylov (JFNK) solver \citep{knoll2004jacobian} with physics-based preconditioning \citep{viallet2016jacobian, goffrey2017benchmarking}.  The \music code uses an adaptive time step, which is identically constrained for all simulations in this work. The MESA code used the OPAL Type II opacities and corresponding equation of state to produce these models; our \music simulations use tables generated from MESA.  In eq. \eqref{momeq}, $\vec{g}$ is the gravitational acceleration, a spherically symmetric vector, consistent with that used in the stellar evolution calculation, and not evolved during the present simulations.

\begin{figure*}[t!]
\begin{center}
\includegraphics[width=0.85\textwidth]{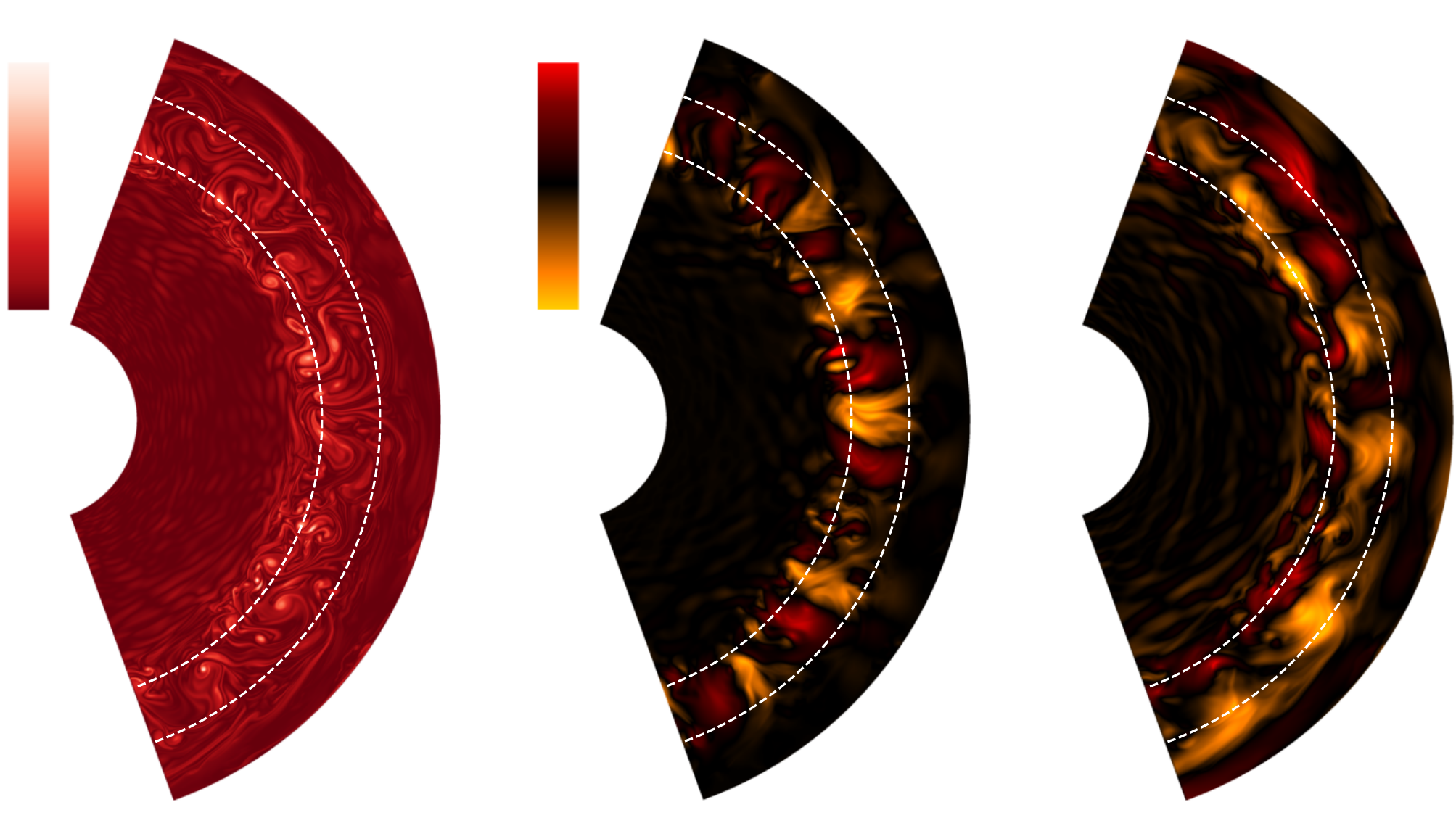}
\caption{Visualizations of (from left to right) vorticity magnitude, radial velocity and angular velocity of the $8 \Msun$ simulation after $30 \tau_\mathsf{conv}$ of steady-state convection. The zero point for vorticity is in dark red in the left panel. In the center and right panels, outwards flows are in red while inwards flows are in yellow; the zero point in velocity is black. The radial velocity ranges from $-9.25 \times 10^5$ cm/s to $9.25 \times 10^5$ cm/s at this instant in time, while the angular velocity ranges from $-1.01 \times 10^6$ cm/s to $1.01 \times 10^6$ cm/s.  Dashed white lines indicate the position of the Schwarzschild boundaries.}
\label{visu}
\end{center}
\end{figure*}

\begin{figure}[h!]
\begin{center}
\includegraphics[width=\columnwidth]{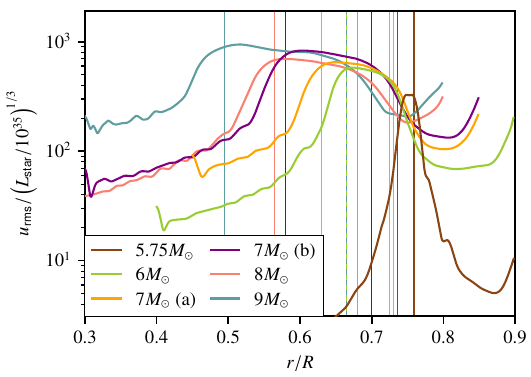}
\caption{Radial profile of the time-averaged RMS velocity scaled by $\left(L_{\mathsf{star}}/10^{35}\right)^{1/3}$ for a range of stellar masses. Vertical lines indicate the Schwarzschild convective boundaries for each star.}
\label{vrms}
\end{center}
\end{figure}

We study the \music simulations described in Table~\ref{simsuma}.
For all of these simulations, the compressible hydrodynamic equations \eqref{densityeq}-\eqref{ieneq} are solved in a spherical shell using spherical coordinates ($r$, $\theta$). Each simulation is two dimensional, and assumes azimuthal symmetry.  In the table, the inner and outer radii of the spherical shell are noted, as well as the upper and lower boundary of the convection zone. The simulation domain includes the interior convection zone and is truncated just below the outer convective envelope (see Fig.~\ref{sketch}).

In Table~\ref{simsuma}, we compare the pressure scale height to the grid spacing at a convective boundary to measure how well the physics around that boundary is resolved.  This results in two resolution criteria for a convective shell: the upper resolution $H^{\mathsf{U}}_{\mathsf{p,CB}}/\Delta r$  and the lower resolution $H^{\mathsf{L}}_{\mathsf{p,CB}}/\Delta r$. These two resolutions are fairly similar.  In each case there are more than 100 grid spaces per pressure scale height, ensuring a high resolution of the overshooting layer.   Each of our simulations has sufficient radial resolution to produce a characteristic radial profile for velocity in 2D, and convergence toward a velocity profile is observed as the grid is increased.  Simulation ceph8 uses a grid of size $n_r \times n_\theta = 1728 \times1024$.

We examine simulations with boundary conditions that maintain the original radial profiles of density and temperature of the 1D stellar evolution model. We hold the energy flux and luminosity constant on the outer radial boundary at values established by the stellar structure.  For an examination of how boundary conditions affect the dynamics in stars, we refer to \citet{pratt2016spherical,vlaykov2022impact}.
Aside from this surface boundary condition, in velocity we impose non-penetrative and stress-free boundary conditions in the radial directions.  The energy flux and luminosity are held constant at the inner radial boundary, at the value of the energy flux at that radius in the one-dimensional stellar evolution calculation.
On both the inner and outer radial boundaries of the spherical shell, we impose a boundary condition on the density that maintains hydrostatic equilibrium by assuming constant internal energy and constant radial acceleration due to gravity in the boundary cells \citep{hsegrimm2015realistic}.   We impose periodicity on all physical quantities at the boundaries in $\theta$.

We define the convective turnover time following \cite{pratt2016spherical}
\begin{eqnarray}
    \tconv (t) = \int_{\mathsf{CZ}}\mathsf{dV} \frac{H_{\mathsf{p}}}{\left|u\right|} \left/ \int_{\mathsf{CZ}}\mathsf{dV} \right. ~,
    \label{tau_conv}
\end{eqnarray}
where $H_{\mathsf{p}}$ is the pressure scale height and $\DPS \left|u\right|=\sqrt{u_r^2+u_\theta^2}$ the magnitude of the velocity computed in 2D from the radial and angular velocities, $u_r$ and $u_\theta$ respectively. The integration covers the convection zone and is volume-weighted using the volume element $\mathsf{V}$. We use the convective turnover time to verify that convection has reached a steady-state, a period where the time-averaged value of the convective turnover time is well-defined and not evolving in time. For each simulation, the time-averaged value of $\tconv$ in the steady state regime is included in Table~\ref{simsuma} alongside its standard deviation. In general, when we describe time-averaged quantities, this refers to a time average taken over the entire time in steady state covered by the simulation. This time span is more than $25 \tconv$ in every case, ensuring enough data to produce well converged statistics.  The convective boundaries do not evolve during the time span of our hydrodynamic simulations: the temperature at the Schwarzschild boundaries varies by less than $0.8 \%$ for all models. This is expected because the Kelvin-Helmholtz timescale of the stars we examine is at least $4$ orders of magnitude longer than the time covered by our simulations.

In our simulations, the maximum amplitude of the convective velocity in the convective shell is approximately proportional to cube-root of luminosity, $\DPS u_{\mathsf{rms}} \propto L_{\mathsf{star}}^{1/3}$ (see Fig.~\ref{vrms}), a result also found by previous works investigating convective shells \citep[e.g.,][]{jones2016idealised, andrassy20203d}, convective cores \citep[e.g.,][]{higl2021calibrating, horst2020fully, baraffe2023study} and convective envelopes \citep[e.g.,][]{baraffe2021two}. The root-mean-squared (RMS) velocities just above and below the convective boundaries (indicated by vertical lines in Fig.~\ref{vrms}) are caused by convective overshooting that reaches into the neighboring stable regions.   Far from the convection zone, the non-zero velocities are due to the propagation of internal gravity waves (IGWs), excited by the convective motions and overshooting plumes. Such waves, as well as the fully convective internal region, can be observed in visualizations of vorticity and radial and angular velocities (see Fig.~\ref{visu}). The present paper only focuses on convection and convective boundary mixing, and the propagation of IGWs will not be investigated any further.

\section{Results \label{sec:results}}

To describe convection in stellar interiors, the concept of a filling factor has been applied widely \citep[e.g.][]{zahn1991convective, canuto1998stellar, dethero2024shape}.A filling factor is a crude measure that reduces the complex flow of stellar convection to a one-dimensional measure of the amount of fluid moving inward (or outward) at a given radius. A large part of the value of a filling factor is as a measure of the asymmetry between inflows and outflows \citep[e.g.][]{zhang1997non,brown2007temperature,dethero2024shape}.

\begin{figure}[t!]
\begin{center}
\includegraphics[width=\columnwidth]{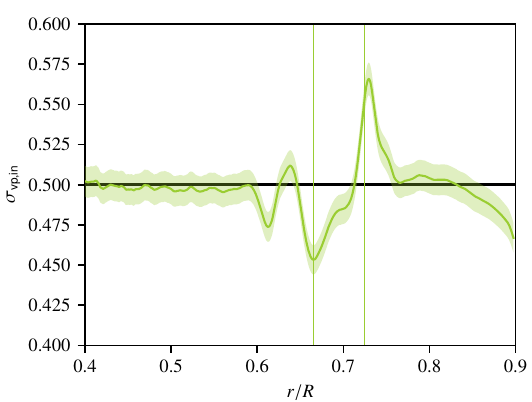}
\caption{Radial profile of the time-averaged volume percentage filling factor of the inward moving plumes $\sigmavp$ vs the star's internal radius, in units of the total stellar radius $R$ for the $6 \Msun$ simulation. The shaded area indicates one standard deviation above and below this averaged line and thin vertical lines indicate the Schwarzschild boundaries.}
\label{ff_vp}
\end{center}
\end{figure}

\begin{figure}[t!]
\begin{center}
\includegraphics[width=\columnwidth]{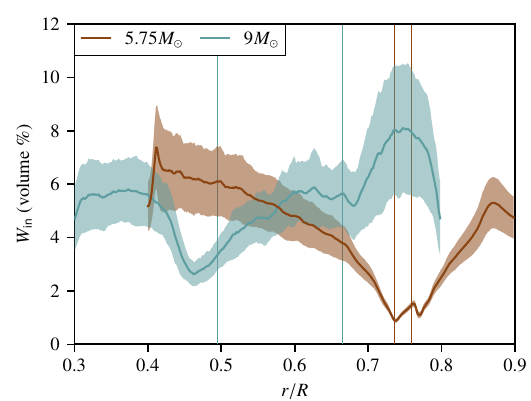}
\caption{Width of inflowing plumes $W_{\mathsf{in}}$ for the $5.75$ and the $9 \Msun$ simulations as a function of the star's radius.  Shaded areas indicate one standard deviation above and below the time-averaged line. Thin vertical lines indicate the radial position of the Schwarzschild boundaries. }
\label{ff_W}
\end{center}
\end{figure}

Several definitions of the filling factor have been proposed in the literature \citep{dethero2024shape}. First, we consider the filling factor based on a simple volume-percentage of inflows at a given radius \citep{schmitt1984overshoot, brummell2002penetration, andrassythesis,kapyla2023convective}. We define the volume-percentage filling factor as the fraction of volume occupied by the inflows
\begin{eqnarray}
    \sigmavp &=&\frac{V^{\mathsf{in}}(r,t)}{V^{\mathsf{in}}(r,t)+V^{\mathsf{out}}(r,t)}~.
    \label{sigma_vp_in}
\end{eqnarray}
Here $V^{\mathsf{in}}(r,t)$ indicates the total volume of grid cells at a given radius, angular position, and instant in time that has an inward velocity. Similarly, $V^{\mathsf{out}}(r,t)$ indicates the total volume of grid cells with an outward velocity.  A value of $\sigmavp=1/2$ indicates symmetry between inflows and outflows, and is expected in the stable radiative regions. In convection zones and overshooting layers, it is possible to obtain highly asymmetric convection patterns, depending on the density stratification of the star in those layers.  The radial profiles of the time-averaged volume-percentage filling factor for inflows remain close to $0.5$ (see Fig.~\ref{ff_vp}); this indicates only a slight asymmetry between inflows and outflows in the convective shells considered in this work. A filling factor $\sigmavp$ less than one half corresponds to inflows that are thinner, faster, and more concentrated than outflows.   For the convective shells that we study, $\sigmavp$ has a distinctive sinusoidal shape within the convectively unstable region; it is highest at the outer boundary of the convective shell, and drops to its lowest value at the lower convective boundary.  This trend is clearly identifiable for all six of the stars, and it is the opposite of what is generally found for outer convective envelopes in other star types \citep{dethero2024shape}. Observational data indicates filling factors less than a half at the solar surface; \music simulations have previously reproduced this observational result, and showed that the filling factor rises to $1/2$ by the convective boundary \citep{dethero2024shape}. Thus, although the convective shells that we consider are in the outer regions of the star, where convective envelopes are commonly studied, the radial trend of asymmetry is more similar to convective cores.  Although statistically significant, this trend is slight, and the convective shells are therefore most similar to laboratory experiments of Boussinesq convection.

A motivation for the concept of the filling factor is to establish a universal diagnostic to quantify the symmetry or the asymmetry of inflows and outflows. However, as pointed out recently by \cite{dethero2024shape}, the filling factor is a number that sums up the width of inflows, and is thus related to a low-order statistic. As an attempt to pursue high-order statistics, \cite{dethero2024shape} proposed to examine the width of inflows directly. The width of inflows (outflows) is defined as the width of a continuous set of cells in the angular direction, at a given radius, which all have a negative (positive) radial velocity. The multi-scale nature of stellar convection can be understood through the radial distribution of plume widths. For large convection zones, an extreme density stratification can lead to a wide range of convective scales. 

In our simulations of relatively thin convective shells, the width of inflows in the convection zones varies only mildly except near the boundaries (see Fig.~\ref{ff_W}).

For example, the time-averaged width of inflows increases from only $1 \%$ of the volume at that radius between the two boundary layers of the $5.75\Msun$ Cepheid, and $3 \%$ for the $9 \Msun$ simulation, ultimately leading to convective rolls of almost the same size in the heart of the convection zone. The overshooting layers both above and below the convection zones are readily identified by minima in the widths of convective structures.  These minima appear to indicate a radial position where plumes defined by radial velocity dissipate to their thinnest extent; the angular resolution of the present simulations would be able to resolve plumes ten times smaller if they occurred.  Our diagnostic continues to calculate widths of radial flows measured further from the convection zone; these are contaminated by the small radial velocities of IGWs, which tend to have larger angular extents.  Any detailed study of such wave dynamics will be undertaken in future work.  The width of inflows in the top overshooting layer is systematically larger than those from the bottom overshooting layers, as indicated by the relative minimum values. Different scales of the convection rolls exist in the two distinct overshooting layers. This is in agreement with the results obtained with the volume-percentage filling factor, i.e inflows are fatter at the top while outflows are fatter at the bottom of the convective shell.

Both the volume-percentage filling factor (Fig.~\ref{ff_vp}) and the width of the inflowing plumes (Fig.~\ref{ff_W}) demonstrate that inside these thin convective shells, apart from the boundary layers, convective flows of approximately the same size dominate.  We do not observe a multi-scale convective flow to the extent  described in larger convective envelopes.  This result can be understood most clearly by looking at the visualization of vorticity in Fig.~\ref{visu}. It shows the presence of convection rolls slightly larger than the width of the convection zone, which overshoot both convective boundaries. Large convective rolls dominate in the convectively unstable region, while thinner and faster inflowing plumes are found close to the Schwarzschild boundaries, consistent with a picture of boundary layer flows.

\begin{figure*}[t]
\begin{center}
\includegraphics[width=\textwidth]{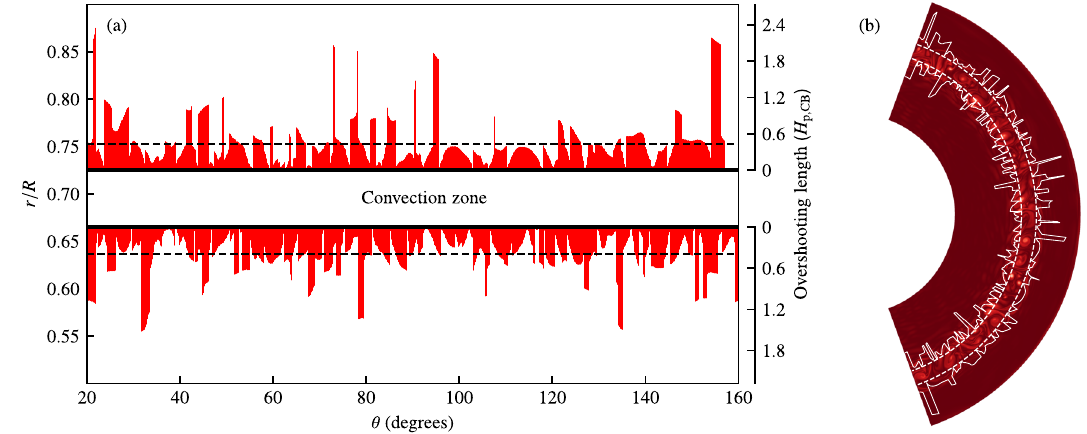}
\caption{(a) Angular structure of the overshooting layer after $114 \tau_\mathsf{conv}$ of steady-state convection in the simulation of the $6\Msun$ Cepheid. The overshooting length in this illustration is determined by zeros of the vertical kinetic flux. The Schwarzschild convective boundaries between the convection zone and the neighboring stable radiative zones are indicated by a solid black line. The stellar radius ($r/R$) is indicated on the left axis, and the overshooting length is given in units of the pressure scale height $H_{\mathsf{p,CB}}$ at the convective boundaries on the right axis. A dashed line indicates the average overshooting length at this time. (b) The same, but plotted over a visualization of vorticity. White dashed lines indicate the Schwarzschild boundaries.}
\label{layer}
\end{center}
\end{figure*}

\subsection{Determination of the extent of the overshooting layer}\label{sec:overshooting}

Convective overshooting is a fundamental phenomenon of stellar physics. Its influence on one-dimensional stellar evolution models is typically accounted for through an overshooting length. 
Determining the extent of convective overshooting requires a method for determining what part of the fluid is inside a convective flow structure and what part is outside.  This is potentially a complex problem because convection drives shear flows and turbulence, and because convective flow structures can be defined both by their large-scale radial motions and by their ability to transport heat. Several definitions have been used to measure the depth of the overshooting layer: the thermal flux, the kinetic energy flux, and the vertical kinetic flux. In this work, we use the vertical kinetic energy flux to define the extent of convective overshooting 
\citep[e.g]{hurlburt1986nonlinear, hurlburt1994penetration, ziegler2003box, rogers2006numerical, tian2009numerical, chan2010overshooting, pratt2017extreme}. The vertical kinetic energy flux $F_\mathsf{k}$ is defined
\begin{equation}
    F_\mathsf{k} = \frac{1}{2} u_r \rho \left| u \right|^2~.
\end{equation}
For convective shells where overshooting can happen both above and below the convective layer, we examine the depth of the upper and lower overshooting layers using the first zero of the vertical kinetic energy flux in the stable radiative zone. This is a method that has been used by \citet{pratt2017extreme, pratt2020comparison, baraffe2021two, vlaykov2022impact,baraffe2023study, dethero2024shape}. 

In our hydrodynamic simulations with no rotation, the overshooting length is statistically independent of the angular position $\theta$. However, convective overshooting varies in $\theta$ at any given instant in time (see Fig.~\ref{layer}(a)). The vertical kinetic flux in the overshooting layer exhibits round shapes over several degrees. These shapes represent large convective rolls protruding from each side of the convection zone that contribute to large overshooting layers (see Fig.\ref{layer}(b)). Instead of dissipating right after the convective boundaries, convective motions that are parts of these convection rolls return into the CZ.  A straightforward approach to evaluate a general overshooting length is to calculate a simple time and space average.  For example, the dashed line in Fig.~\ref{layer}(a) indicates the average in $\theta$ of the overshooting length at the particular instant in time displayed. Despite its simplicity, this straightforward approach hides physically important features of convective overshooting and masks contributions of the convective motions that penetrate further than the average overshooting plume.

\begin{figure}[t]
\begin{center}
\includegraphics[width=\columnwidth]{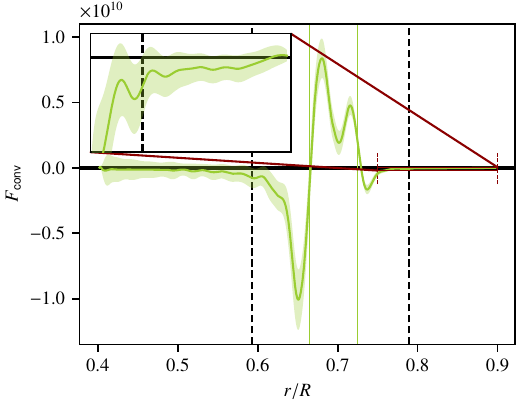}
\caption{Radial profiles of the convective flux for the $6\Msun$ Cepheid. The shaded area indicates one standard deviation around the mean value.  Thin vertical lines indicate the convective boundaries determined by the Schwarzschild criterion.  Black dashed lines indicates one pressure scale height above and below the convection zone.  The inset shows a zoomed in region centered on the red box.}
\label{ff_F_zoom}
\end{center}
\end{figure}

A method used historically \citep{browning2004simulations, brun2011modeling} to calculate the radial extent of the overshooting layer uses the time-averaged radial profile of the convective flux.  We define the convective flux following \cite{freytag1996hydrodynamical, pratt2016spherical}
\begin{equation}
    F_\mathsf{conv} = \left\langle H \rho u_r\right\rangle -\left\langle H \right\rangle \left\langle \rho u_r \right\rangle~,
\end{equation}
where $\DPS \langle \square \rangle$ denotes a time average, and the enthalpy is $\DPS H= e + p/\rho$. From this flux, the overshooting layer is defined using the negative peak in the convective flux that occurs around the convective boundary; the radial point in the radiative zone where this peak becomes negligible is used to define the extent of the overshooting layer.  This method of calculating the overshooting depth has two disadvantages.  Firstly, the volume-averaged profile of the convective flux requires a long-time average to converge to a smooth profile (see Fig.~\ref{ff_F_zoom}).  The convective flux thus provides a useful approximation of the width of the overshooting layer, but cannot be regarded as a precise measurement.  Secondly, in the overshooting layer and radiative zone, the time-averaged profile of the convective flux is contaminated by small-scale oscillations at the point where it would become negligible; these may be the signature of internal gravity waves. As a consequence, some judgment must be exercized to estimate this point. We use the convective flux estimate of the overshooting layer as a reasonable check for the more accurate statistical methods that we describe next.  Overall, the convective flux shows that the size of the overshooting layers is of the same order of magnitude as the size of the convection zone. Approximate values of the overshooting depth for the upper and lower overshooting layers, calculated on the basis of the convective flux profile are included in Table~\ref{ov_comp}, and are denoted $\overline{\DPS \ell_\mathsf{ov}}$. A superscript $\mathsf{U}$ or $\mathsf{L}$ differentiates the upper and lower overshooting layers. 

\begin{figure}[t!]
\begin{center}
\includegraphics[width=\columnwidth]{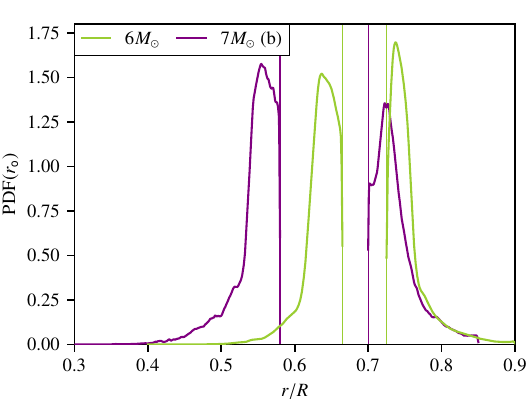}
\caption{PDF of $\ro$, the extent of all overshooting plumes, for the $6$ and the $7 \Msun$(b) Cepheid. Thin vertical lines indicate the convective boundaries determined by the Schwarzschild criterion.}
\label{pdf_all}
\end{center}
\end{figure}

A more precise method for defining the extent of the overshooting layer is to use instantaneous values of the vertical kinetic energy flux. We define the position $\ro$ as the position of the first zero of the vertical kinetic energy flux. The probability density functions (PDF) of the overshooting extent of all plumes for the $6$ and the $7 \Msun$(b) are shown in Fig.~\ref{pdf_all}; for the other simulations, the PDF($\ro$) have a similar shape. As observed by \cite{pratt2017extreme} in the case of large convective envelopes, these PDFs have a single tail that is heavier than a Gaussian distribution, and are asymmetrical because plumes do not stop inside the convection zone. These heavy tails suggest that strong plumes reach deeply in the radiative zone. A straightforward average of $\ro - R_{\mathsf{CB}}$ is therefore not sufficient to determine an overshooting length to describe how convective overshooting affects the stable radiative layer. If a distribution is Gaussian, it can be fully characterized by a mean and a standard deviation. However, for a non-Gaussian distribution like the ones produced here, using the average eliminates the most significant part of this intermittent fluid process. In this section, we first seek to apply the strategy of using extreme value theory (EVT) to evaluate an overshooting length both above and below the convective shell.

\subsection{Application of the EVT}\label{subsec:overshooting_lengths}

\begin{figure}[t!]
\begin{center}
\includegraphics[width=\columnwidth]{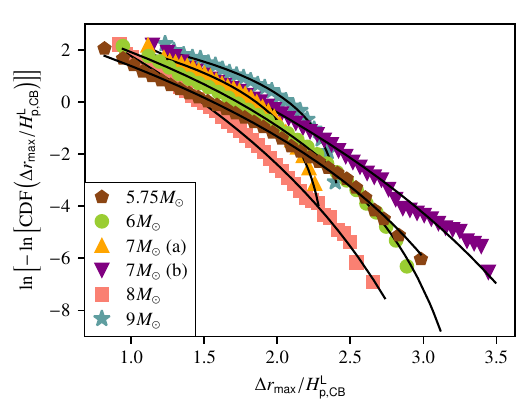}
\caption{Natural logarithm of the negative natural logarithm of the cumulative distribution functions of maximal overshooting length, $\Delta r_{\mathsf{max}}$ for each simulation below the convective shell. Black lines indicate the best fit with the Weibull distribution, defined in eq.~\eqref{weibull}.} 
\label{cdf_max}
\end{center}
\end{figure}

To describe the influence of the most energetic overshooting plumes which penetrate further in the RZ and are responsible for the heavy tail in the PDF($\ro$), we first follow the approach developed by \cite{pratt2017extreme} using the generalized extreme value distribution (GEVD) \citep{castillo2005extreme}. We define the maximal overshooting depth $\rmax$ to be the lowest (highest) position in the radiative zone below (above) the convective shell that is reached at a given time. Below the convective shell, $\rmax$ reads 
\begin{equation}
    \rmax(t) = \mathrm{min}_\theta (\ro(\theta,t)) ~,
\end{equation}
and above
\begin{equation}
    \rmax(t) = \mathrm{max}_\theta (\ro(\theta,t))~.
\end{equation}
The maximal overshooting length $\Delta \rmax$ is then the difference between $\rmax$ and the convective boundary $R_{\mathsf{CB}}$:
\begin{equation}
    \Delta \rmax(t) = \left| \rmax(t) - R_\mathsf{CB} \right|~.
\end{equation}
In the following, $\Delta \rmax$ will be made dimensionless with the pressure scale height at the convective boundary, $\DPS H_\mathsf{p,CB}$.

\subsubsection{Below the convection zone}

Applying EVT, the cumulative distribution function (CDF) of the maximal overshooting length should follow a Weibull distribution:
\begin{equation}
    F(\Delta \rmax)=\exp\left(-\left(1+\alpha\left(\frac{\Delta \rmax - \mu}{\lambda}\right)\right)^{-1/\alpha}\right)~,
    \label{weibull}
\end{equation}
where $\alpha<0$ is the shape parameter, $\lambda$ is the scale parameter and $\mu$ is the location parameter. This distribution provides a good fit to the maximum overshooting data generated by simulations of convective envelopes.  \cite{pratt2017extreme} demonstrated that the association of the overshooting length with the location parameter of the fitted PDF accounts for the effects of the strongest plumes in the stable radiative zones. For overshooting below a convective shell, the CDFs of the maximal penetration length below the convective shell are also fit reasonably well by a Weibull distribution (see Fig.~\ref{cdf_max}). The downward curvature of the CDFs shows that they also follow a Weibull distribution for the current simulations. For each simulation, our EVT analysis provides a maximal width, denoted $\ell^\mathsf{max}_{\mathsf{ov}}$, and derived from the location parameter. These values are summarized in Table~\ref{ov_comp}. The overshooting length we extract from the GEVD fit are significantly larger than what is obtained from a straightforward averaged as done in Fig.~\ref{layer}(a), in agreement with the conclusions of \cite{pratt2017extreme}.

\subsubsection{Above the convection zone: convective boundary mixing throughout the radiative layer}

As explained in Section \ref{sec:sim}, the simulation domain has been truncated at a point just below the outer convective envelope. In the our simulations, this presents a difficulty because plumes overshooting above the convective shell sometimes reach the outer boundary of the simulation; this is rarely the case below the convective shell. Because sometimes plumes reach the outer simulation boudary, maximal-in-time plumes often reach the simulation boundary.
Therefore, the EVT approach produces an overshooting length equal to the depth of the radiative layer for all of the stars we study except for ceph5, The simulation ceph5 represents the thinnest convective shell; the mass and luminosity of this star are lower than for the other stars we study, resulting in lower velocities at the convective boundary. Therefore, the overshooting plumes stop closer to the CZ, and the RZ is larger than the full overshooting layer.

However, the majority of the stars we examine have higher mass and luminosity; the convective motions reach further in the RZ. 
For $M>6\Msun$, the EVT analysis show that the extreme events often hit the outer boundary of the spherical shell.  The shape of the PDF($\Delta \rmax$) approaches a $\delta$ distribution; a $\delta$ distribution would be achieved if, at any time, there is at least one plume that hits the upper boundary of the domain.  A radiative zone that is fully dominated by overshooting plumes has been described in recent work by \cite{guo2021convective}, who studied A-type stars.  The EVT analysis confirms overshooting throughout the RZ. However, it is unable to provide us with a detailed picture of of the upper overshooting layer that can be compared with the lower overshooting layer.

\begin{figure}[t!]
\begin{center}
\includegraphics[width=\columnwidth]{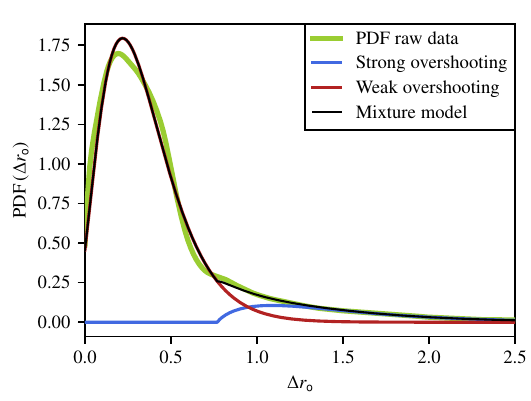}
\caption{PDF($\Delta \ro$) of a $6 \Msun$ Cepheid above the convective shell and best fit with the mixture distribution.}
\label{fit_mixture}
\end{center}
\end{figure}

\begin{figure*}[t!]
\begin{center}
\includegraphics[width=\textwidth]{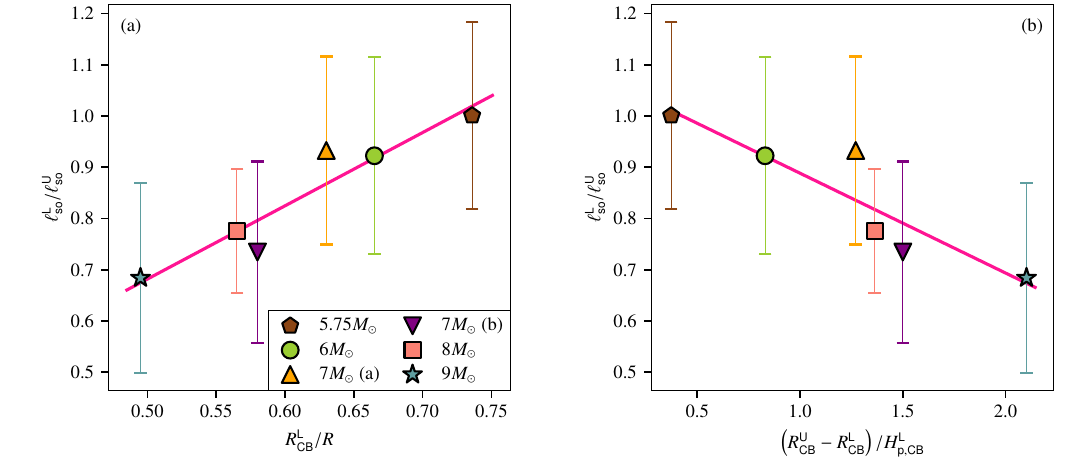}
\caption{Ratio of the extent of the lower overshooting layer ($\ell_\mathsf{so}^\mathsf{L}$) over the extent of the upper overshooting layer ($\ell_\mathsf{so}^\mathsf{U}$) given by the mixture model, (a) vs. the depth of the convective shell, measured as $R_{\mathsf{CB}}^\mathsf{L}/R$, and (b) vs. the width of the convective shell measured as $\left(R_{\mathsf{CB}}^\mathsf{U} - R_{\mathsf{CB}}^\mathsf{L}\right)/H^\mathsf{L}_\mathsf{p,CB}$. Error bars indicate one standard deviation around the mean value. A pink line indicates a linear regression fit.}
\label{ratio}
\end{center}
\end{figure*}

\begin{table*}[!t]
\begin{center}
\caption{Characteristic length scales of the overshooting layers obtained with the radial profiles of the convective flux, the EVT and with the mixture model.  
 \label{ov_comp}
 }
\begin{tabular}{lcccccccccccccccccccccccccccc}
                      & &\textbf{Convective flux}& &\textbf{Extreme value theory} & &\multicolumn{3}{c}{\textbf{Mixture model}} 
\\ \hline \hline
                     
                     &
                     &$\left(\overline{\ell^\mathsf{L}_{\mathsf{ov}}}, \overline{\ell^\mathsf{U}_{\mathsf{ov}}}\right)$& &$\left(\ell^\mathsf{L, max}_{\mathsf{ov}}, \ell^\mathsf{U,max}_{\mathsf{ov}}\right)$& & $\left({\ell^\mathsf{L}_{\mathsf{so}}}, {\ell^\mathsf{U}_{\mathsf{so}}}\right)$& &                    $\left({\ell}^\mathsf{L,90\%}_{\mathsf{so}},{\ell}^\mathsf{U,90\%}_{\mathsf{so}}\right)$&
\\ \hline
ceph5       & & $(0.6,0.6)$ & &$ (1.56 , 1.66)  $   & &    $ (0.47 \pm 0.07, 0.47 \pm 0.29)$ & & $(0.59,1.06)$
\\ \hline
ceph6       & & $(1.0,1.1)$ & & $ (1.73, \mathit{2.41})  $   & &    $ (1.37 \pm 0.33, 1.49 \pm 0.54)$ & & $(2.02,2.53)$
\\ \hline
ceph7a      & & $(1.1,1.6)$ & & $ (1.86, 2.84)  $   & &    $ (1.21 \pm 0.35, 1.30 \pm 0.47)$ & & $(1.89,2.19)$
\\ \hline
ceph7b      & & $(1.2,1.9)$ & &$ (1.87, 3.23)  $   & &    $ (1.29 \pm 0.37, 1.76 \pm 0.61)$ & & $(2.02,2.92)$        
\\ \hline
ceph8       & & $(1.0,1.3)$ & &$ (1.48, 2.60)  $   & &    $ (1.31 \pm 0.27, 1.69 \pm 0.49)$ & & $(1.83,2.60)$      
\\ \hline
ceph9       & & $(1.2,1.7)$ & &$ (2.08, 3.36)  $   & &    $ (1.43 \pm 0.37, 2.14 \pm 0.74)$ & & $(2.15,3.51)$
\\ \hline

\end{tabular}
\end{center}
\begin{tablenotes}
    \item[] \textbf{Notes:} $\DPS \overline{\ell^\mathsf{L}_\mathsf{ov}}$ $\DPS \left(\overline{\ell^\mathsf{U}_\mathsf{ov}}\right)$ is a width of the lower (upper) overshooting layer defined using the convective flux that is first averaged over the angular direction and then averaged in time. $\DPS \ell^\mathsf{L, max}_\mathsf{ov}$ $\DPS \left(\ell^\mathsf{U, max}_\mathsf{ov}\right)$ is a length defined by the EVT analysis, which provides a maximal width of the lower (upper) overshooting layer. For simulations ceph7a, ceph7b, ceph8 and ceph9, the value of the EVT for the upper overshooting layer is the full width of the radiative layer. The value in italics for the $6\Msun$ Cepheid has been obtained from the incomplete PDF($\Delta \rmax$) and therefore associated with a higher uncertainty. $\DPS {\ell}^\mathsf{L}_\mathsf{so}$ $\DPS \left( {\ell}^\mathsf{U}_\mathsf{so}\right)$ is a characteristic length-scale associated with the lower (upper) overshooting layer, defined by the mean of the SO Gamma distribution.  The standard deviation of the SO Gamma distribution is indicated as an error on the mean. $\DPS {\ell}^\mathsf{L,90\%}_{\mathsf{so}}$ $\left({\ell}^\mathsf{U,90\%}_{\mathsf{so}}\right)$ is the distance at which $90\%$ of the plumes have stopped below (above) the CZ, defined as the $90^\mathrm{th}$ percentile of the SO Gamma distribution. All values are given in units of the pressure scale height at the convective boundary, $H_\mathsf{p,CB}$.
\end{tablenotes}
\end{table*}

\subsection{Description of the PDF($\ro$) using a mixture distribution of Gamma functions}

The EVT analysis of overshooting was developed by \cite{pratt2017extreme} to describe the intermittency of overshooting and the influence of the most energetic overshooting events;  the intermittency of this process was underestimated when a simple average is used because of the non-Gaussian nature of the PDF($\ro$). This approach provides us with a maximal width for the overshooting layers. However, we need a more detailed picture of the upper overshooting layer to compare with overshooting below the convective shell. Here, we propose a new way to characterize the intermittency of overshooting events. Strongly overshooting plumes are responsible for the heavy tail in the PDF($\ro$). To model the non-Gaussian nature of the distribution of all plumes, we propose to separate the contribution of the strongest plumes from that of the plumes that simply make it just beyond the boundary of the convective instability. These weakest overshooting plumes play a negligible role in the enhanced diffusion throughout the radiative zone. The PDF($\ro$) reveals that most overshooting plumes stop immediately after the convective boundary (see in Fig.~\ref{pdf_all}). Taking the average value of these non-Gaussian distributions therefore overestimates the influence of the weakest plumes and underestimates how the strongest overshooting events enhance the mixing mechanisms in the RZ. 

\subsubsection{Description of the mixture model}

In an effort to separate the contributions of the weak overshooting from the strong one, we therefore describe these distributions with a mixture model of two Gamma distributions. 
First, we consider the strictly positive random variable $\DPS \Delta \ro$, defined as
\begin{eqnarray}
    \DPS \Delta \ro = \left| \frac{\ro - R_\mathsf{CB}}{H_\mathsf{p,CB}} \right| ~.
\end{eqnarray}
The choice of the Gamma distribution is motivated by the fact that it is the maximum entropy probability distribution for a strictly positive random variable, given its mean value and the mean value of its logarithm \citep{park2009maximum}. Essentially, the Gamma distribution is the most general distribution that describes a strictly positive random variable with a positive skewness. 

The probability density function of a Gamma distribution has the form
\begin{eqnarray}
    f(x;\alpha, \lambda, \mu) = \frac{1}{\Gamma(\alpha) \lambda^\alpha}(x-\mu)^{\alpha - 1} \exp\left(\frac{-(x-\mu)}{\lambda} \right)~.
    \label{PDF_gamma}
\end{eqnarray}
Here $x$ is a coordinate that will be substituted for the variable $\Delta \ro$ in our calculations. The cumulative distribution function of the Gamma distribution is
\begin{eqnarray}
    F(x;\alpha, \lambda, \mu) = \frac{1}{\Gamma(\alpha)}\gamma \left(\alpha, \frac{x-\mu}{\lambda} \right)~,
\end{eqnarray}
where $\alpha$, $\lambda$ and $\mu$ are parameters respectively known as the shape, the scale, and the location parameter. The lower incomplete Gamma function $\gamma$ is defined for every complex number $z$ with a strictly positive real part
\begin{equation}
    \gamma (z,x) = \int_0^x t^{z-1} e^{-t} \mathrm{d}t~.
\end{equation}
Here $\Gamma$ is the Gamma function, defined as the limit of $\gamma$ when $x$ tends to infinity. The mixture model that we apply is composed of two Gamma distributions. The first describes the peak of the PDF($\Delta \ro$), and will be referred to as the weak overshooting (WO) distribution ($f_\mathsf{wo}$). The second describes the contribution of the heavy tail of the PDF($\Delta \ro$), which corresponds to the most energetic events. This second distribution will be called the strong overshooting (SO) distribution ($f_\mathsf{so}$). The PDF of the mixture model then reads
\begin{equation}
    f_\mathsf{mix}(\Delta \ro) = w_\mathsf{wo} f_\mathsf{wo}(\Delta \ro) + w_\mathsf{so} f_\mathsf{so}(\Delta \ro)~.
    \label{mixture_model}
\end{equation}
Here $w_\mathsf{wo}$ and $w_\mathsf{so}$ are weights of the WO and SO distributions, respectively, such that $w_\mathsf{wo}$,  $w_\mathsf{so} \ge 0$ and $w_\mathsf{wo} + w_\mathsf{so} = 1$.

We have written a fitting program which first calculates the CDF($\Delta \ro$) of the data set and defines the mixture model given in eq.~\eqref{PDF_gamma} and eq.~\eqref{mixture_model} using $w_\mathsf{so}=1-w_\mathsf{wo}$. Our mixture model has seven free parameters: the location, shape and scale parameters for each of the two Gamma distributions and the weight. We then employ a Levenberg–Marquardt algorithm \citep{levenberg1944method, marquardt1963algorithm} to find the best fit for the data. This is achieved using a non-linear least squares method to minimize the residual and find the optimal values for the $7$ parameters of the mixture distribution. The Levenberg-Marquardt algorithm is an iterative procedure which requires an initial guess. We test the robustness of this algorithm for the current problem by checking that a range of different initial guesses converge to the same solution. Application of this mixture distribution to the PDF($\Delta \ro$) above the CZ of the $6 \Msun$ star is shown in Fig.~\ref{fit_mixture}. The mixture model produces a useful splitting of the original PDF into two Gamma distributions. In each case, the WO distribution corresponds to approximately $85\%$ to $95 \%$ of the overall PDF, confirming that the weakest overshooting events are more frequent than the more energetic ones.  Because the WO distribution includes  the measured overshooting of convective rolls that protrude from each side of the CZ, it makes physical sense that contains a much larger portion of the probability space than the SO distribution. The mixture model therefore allows us to filter the contribution of the weak overshooting plumes and model exclusively the events that most enhance the mixing mechanisms in the RZ.

\subsubsection{Estimation of the overshooting length with the mixture model and comparison with the EVT}

The mixture model uses the heavy tail of the PDF of all plumes to produce a full distribution for the strong overshooting events. The mean value that we extract from this distribution represents an average depth reached by strong overshooting plumes inside the RZ; we denoted this depth by $\ell_\mathsf{so}$.  It represents a characteristic length scale associated with enhanced diffusion within the RZ. We find a relatively large standard deviation around the expected value.  These values are summarized in Table \ref{ov_comp}. Using both the mixture model and the EVT provides a detailed picture of the strong overshooting layer. Indeed, the EVT provides a maximal width of the overshooting layer.

Both the EVT and the mixture model yield length scales for overshooting that are larger than what is usually found for either convective envelopes or convective cores. These large length scales are consistent with the approximate width of the overshooting layer given by the averaged profile of the convective flux, as well as visualizations. For example, the visualization of vorticity in Fig.~\ref{visu} clearly shows that convective rolls can be larger than the width of the CZ itself and can protrude on either side. Convective plumes are projected ballistically away from the convective boundaries with high energy due to the centrifugal force of the rotating vortices. These high energies appear to be increased by the opposite rotation of side-by-side convection rolls.  This regime of convection differs significantly from large convective cores or envelopes.

In addition to the mean value, Table \ref{ov_comp} includes the quantity ${\ell}^\mathsf{L,90\%}_{\mathsf{so}}$ $\left({\ell}^\mathsf{U,90\%}_{\mathsf{so}}\right)$.  This length scale is defined as the $90^\mathrm{th}$ percentile of the SO Gamma distribution, the distance from the convection boundary in the lower (upper) overshooting layers, where $90\%$ of the strongly overshooting plumes have already stopped. The $90\%$ length scale is comparable to the maximal width of the overshooting layer provided by the EVT, and thus provides a link between these two statistical analysis methods.  The $90\%$ SO length scale reinforces the idea that the upper overshooting layer fills its radiative zone; it also provides a picture of how many plumes reach the outer convective envelope.  $10\%$ of the strongly overshooting plumes move beyond this length, and the strongly overshooting plumes are no more than $15\%$ of the full distribution of overshooting plumes; thus an upper estimate is that $1.5\%$ of overshooting plumes can reach, and be mixed directly into the outer convective envelope of any of these Cepheids.  By providing the $90\%$ length scale alongside the mean value of the SO distribution, we indicate a rough rate for which the number of overshooting plumes are able to mix each radial point within the radiative zone.

The ratio of the overshooting lengths for convective shells is uncertain, and is a measurement that can directly benefit stellar evolution calculations.  Using 3D hydrodynamic simulations of massive stars, \cite{cristini20173d, cristini2019dependence} and \cite{rizzuti2022realistic} found the ratio of lower overshooting length to upper overshooting length to be $0.2$. For thin and shallow shells, the $k-\omega$ model of \cite{guo2021convective} predicted a ratio of $0.5$.  These two numbers mark out a wide range of values.

To understand where the convective shells in Cepheids fall, we describe the amount of convective boundary mixing both above and below the shell using our mixture model.  To describe this, we examine the ratio of the lower overshooting length to the upper overshooting length, $\DPS \ell_\mathsf{so}^\mathsf{L} / \ell_\mathsf{so}^\mathsf{U}$.  This ratio is close to $1$ for shallow convective shells.  We use the  position of the lower convective boundary to determine how deep the convective shell lies within the star.
The ratio of overshooting lengths decreases as the shell is located closer to the core of the star (see Fig.~\ref{ratio}(a)). 

The depth of the shell within the star also correlates with its width, measured as $\left(R_{\mathsf{CB}}^\mathsf{U} - R_{\mathsf{CB}}^\mathsf{L}\right)/H^\mathsf{L}_\mathsf{p,CB}$.  In our study, the deepest convective shell is more than four times wider than the smallest, by this measure.
The ratio of overshooting lengths decreases as the shell becomes thicker (see Fig~\ref{ratio}(b)). This makes sense because the density difference above and below a thin, internal convection shell is small; the difference between the RZ above and below the shell is also small.  Moreover, for such thin shells, we observe that the CZ is filled with a line of convection rolls that fills the CZ, producing a flow that is dominated by a single length scale. We thus expect the velocities of the convective rolls to be approximately the same at the bottom and the top of the CZ. When the shell is thicker, the overshooting ratio is smaller, due to the density difference between the lower and upper RZs.

The $8 \Msun$ Cepheid has a deeper shell than the $7 \Msun$ (b), but it is also thinner, demonstrating the combination of these effects. However, the small number of MESA models we selected for our hydrodynamic simulations differ more than just the mass and luminosity of the star.  Because this work includes only high resolution simulations of 6 stars, it is unclear to what extent the depth or width of the shell influences the overshooting ratio independently.  A more exhaustive parametric study would be required to fully characterize these thin and shallow convection zones and confirm these trends.

\subsubsection{Determination of a diffusion coefficient for convection} 

\begin{figure}[t!]
\begin{center}
\includegraphics[width=\columnwidth]{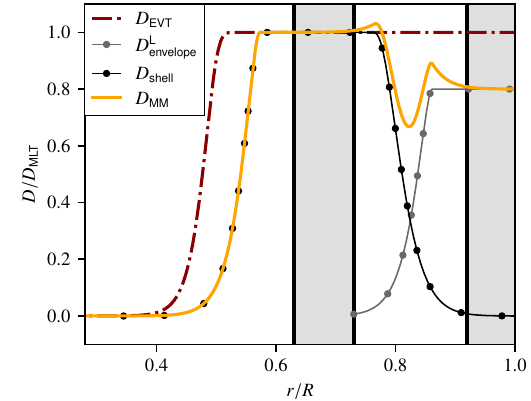}
\caption{Diffusion coefficient for the $7\Msun$ (a) Cepheid. Heavy black lines indicate the convective boundaries of both the convective shell and the outer convection zone. The thin black dotted line is the diffusion coefficient of the convective shell based on the mixture model proposed in eq.~\eqref{diff_coef}, and the red dashed line is the diffusion coefficient based on the EVT analysis (eq.~\eqref{diff_evt}). A gray line extrapolates our model to show a possible diffusion coefficient for the outer convection zone.  The solid yellow line is a model for mixing in the radiative zone that experiences overshooting from both above and below, based on eq~\eqref{diff_coef_full}. In this example, $D_\mathsf{MLT}^\mathsf{envelope} = 0.8~D_\mathsf{MLT}$.}
\label{diff_coef_fig}
\end{center}
\end{figure}

Based on these results, we propose a new form of the diffusion coefficient used to model convective overshooting, designed for the complex situation of a radiative zone between two convective layers.  \cite{pratt2017extreme} demonstrated that the diffusion coefficient can be expressed using probability distributions of overshooting plumes with the simple formula
\begin{equation}
    D_\mathsf{EVT}(r) = D_\mathsf{MLT} \left(1 - F(r) \right)~. 
    \label{diff_evt}
\end{equation}
In this expression, $F(r)$ is the CDF as a function of the star's internal radius; in \cite{pratt2017extreme} the form for the CDF was provided by EVT.  The diffusion coefficient given by MLT is evaluated at the convective boundary, and defined
$\DPS D_\mathsf{MLT} = 1/3 ~ L_\mathsf{MLT} ~ v_\mathsf{MLT}$
; the MLT length scale $L_\mathsf{MLT}$ is proportional to the pressure scale height.  Using the CDF of the mixture model, we propose the following form for mixing enhanced by convective overshooting
\begin{eqnarray}\centering\begin{split} \label{diff_coef}
    D_\mathsf{shell} =& D_\mathsf{MLT} \left[1 -  \frac{1}{\Gamma\left(\alpha^\mathsf{L}_\mathsf{so}\right)}\gamma \left(\alpha^\mathsf{L}_\mathsf{so}, \frac{\left(R^\mathsf{L}_\mathsf{CB}-r\right)/R-\mu^\mathsf{L}_\mathsf{so}}{\lambda^\mathsf{L}_\mathsf{so}} \right) \right.\\
    & \left. - \frac{1}{\Gamma\left(\alpha^\mathsf{U}_\mathsf{so}\right)}\gamma \left(\alpha^\mathsf{U}_\mathsf{so}, \frac{\left(r-R^\mathsf{U}_\mathsf{CB}\right)/R-\mu^\mathsf{U}_\mathsf{so}}{\lambda^\mathsf{U}_\mathsf{so}} \right) \right]~.
\end{split}
\end{eqnarray}
The location parameter $\DPS \mu^\mathsf{U(L)}_\mathsf{so}$, the shape parameter $\DPS \alpha^\mathsf{U(L)}_\mathsf{so}$ and the scale parameter $\DPS \lambda^\mathsf{U(L)}_\mathsf{so}$ of  the SO distribution of the upper (lower) overshooting layer may be estimated from the simulations we performed in this work and give a range (in units of the pressure scale height at the convective boundary)

\begin{subequations}
\begin{align}
&\mu^\mathsf{U}_\mathsf{so} \approx 0.5 - 0.9 H_\mathsf{p,CB}^\mathsf{U} ~, \quad    &\mu^\mathsf{L}_\mathsf{so} \approx 0.7 - 1.0 H_\mathsf{p,CB}^\mathsf{L} ~,  \tag{\theequation a,b} \\
&\alpha^\mathsf{U}_\mathsf{so} \approx 2.0 - 5.0 ~,  \quad &\alpha^\mathsf{L}_\mathsf{so} \approx 1.5 - 3.0~, \tag{\theequation c,d} \\
&\lambda^\mathsf{U}_\mathsf{so} \approx 0.2 - 0.4 H_\mathsf{p,CB}^\mathsf{U}~,  \quad &\lambda^\mathsf{L}_\mathsf{so} \approx 0.15 - 0.25 H_\mathsf{p,CB}^\mathsf{L}~.\tag{\theequation e,f}
\end{align}
\end{subequations}
Unlike the EVT analysis, neither of the location parameters used in the mixture model are directly equivalent to an overshooting length. This model for the diffusion coefficient predicts a smaller overshooting layer than the EVT, by definition (see Fig.~\ref{diff_coef_fig}). However, rather than filtering out all plumes except the one that penetrates the deepest, the new mixture model method produces a full distribution of overshooting plumes that matter, by filtering out only the weak overshooting plumes that do not. It assumes full mixing in this weak overshooting region. This allows for the amount of overshooting within a radiative layer between two convection zones to be modeled in a more detailed way.

Our simulations did not include the outer convective envelope.  However, our overshooting diffusion coefficient based on the mixture model allows us to consider the combined effects of the convective shell and the convective envelope on the radiative layer between them. The EVT analysis predicts full mixing between these two convective layers, creating an outer envelope that is proportionally at least large as the current sun for these Cepheids. Our new mixture model in eq.~\eqref{diff_coef} provides a detailed picture of how such overshooting layers could overlap.  We assume that, for the outer convective envelope, the mixture model would produce a diffusion coefficient of the same form as for the lower overshooting layer of the convective shell, taking into account the different pressure scale height. We propose the following form of the diffusion coefficient for the full star, denoted $\DPS D_\mathsf{MM}$
\begin{equation}
    D_\mathsf{MM}  = D_\mathsf{shell} + D^\mathsf{L}_\mathsf{envelope}~, 
\label{diff_coef_full}
\end{equation}
where the additional diffusion coefficient is defined
\begin{equation}
    D^\mathsf{L}_\mathsf{envelope} = D_\mathsf{MLT}^\mathsf{envelope} \left[1 -  \frac{1}{\Gamma\left(\alpha^\mathsf{E}_\mathsf{so}\right)}\gamma \left(\alpha^\mathsf{E}_\mathsf{so}, \frac{\left(R^\mathsf{E}_\mathsf{CB}-r\right)/R-\mu^\mathsf{E}_\mathsf{so}}{\lambda^\mathsf{E}_\mathsf{so}} \right) \right]~.
\label{diff_coef_envelope}
\end{equation}
For the purposes of exploring mixing throughout this radiative zone, we assume, for example, $\DPS \mu^\mathsf{E}_\mathsf{so}= \mu^\mathsf{L}_\mathsf{so}$, $\DPS \alpha^\mathsf{E}_\mathsf{so}= \alpha^\mathsf{L}_\mathsf{so}$ and $\DPS \lambda^\mathsf{E}_\mathsf{so}= \lambda^\mathsf{L}_\mathsf{so}$ and that $D_\mathsf{MLT}^\mathsf{envelope} = 0.8~D_\mathsf{MLT}$. 

With these parameters, the mixing in the radiative zone between two convective layers can be calculated 
(an example is shown in Fig.~\ref{diff_coef_fig}).
In the figure, the two overshooting layers overlap. Close to the upper convective boundary, we observe that $\DPS D_\mathsf{MM}/D_\mathsf{MLT}$ can be greater than $1.0$ in some regions.  This finding introduces the idea that ``super-mixing layer'' could exist between two convection zones that are relatively close to each other; such super-mixing layers could have more intense mixing than in the neighboring convection zones.  This idea is consistent with the concept of diffusion that is enhanced by the local properties of convection. Hydrodynamic simulations with high radial resolution, which contain both convective layers would be required to confirm the existence of these ``super-mixing layer'' and more fully explore the statistics necessary for this overshooting model.

\section{Summary and conclusions \label{secconc}}

We have studied the internal structure of intermediate-mass stars that lie in the instability strip, and therefore are Cepheid variable stars. We produced a range of stellar models using MESA and documented the presence of a thin, shallow interior convective shell as the star crosses the instability strip. Such inner convective shells have been studied only rarely using multidimensional hydrodynamic simulations.  We have performed hydrodynamic simulations in spherical shells that cover 40\% to 60\% of the stellar radius, for stellar structures corresponding to masses in the range of $5.75$ to $9\Msun$.  We have studied stellar convection and convective boundary mixing in these simulations using statistical methods.  The statistical results that we produce are based on data sets that cover long periods of stellar convection for each star. 

The volume-percentage filling factor of inflowing plumes reveals only a slight asymmetry between inflows and outflows in these convective shells. The filling factor has a sinusoidal shape for each simulation, indicating larger inflows in the upper half of the convective shell while outflows are bigger in the lower half.  This is opposite to the general picture seen in convective envelopes, where inflows are thicker at the bottom of the convection zone.  Although this trend is significant, the filling factor remains relatively close to $0.5$, and the sinusoidal shape has an amplitude that makes the filling factor deviate away from perfectly symmetric inflows and outflows by approximately $\sim 10\%$. These results are in agreement with the analysis of the average width of inflows. In the unstable region, large convection rolls that fill the width of the layer dominate the flow, and we observe a nearly single-scale convective structure. However, boundary layer flows are strong, and closer to the Schwarzschild boundaries we observe  thinner inflows. In both of the overshooting layers, the widths of plumes reach a distinct minimum.

To define the extent of the overshooting layers, we have used the vertical kinetic energy flux. Historically, a simple time and spatial average has often been used to determine the depth of the penetration layer. However, as \cite{pratt2017extreme} observed for pre-main sequence stars with large convective envelopes, in the stars we study, the PDFs of all penetrating plumes are non-Gaussian. Two distinct layers can be identified in the radiative zone: a shallow layer in which the weakest convective plumes overshoot frequently, and a deeper layer where strong overshooting plumes overshoot more rarely.  The weak overshooting layer is characterized by the presence of convection rolls that exceed the convective boundaries before returning back in the CZ, an unusual regime of stellar convection that differs from both convective cores and envelopes. The strong overshooting plumes are responsible for the enhanced diffusion mechanisms in the radiative zone. We therefore applied the extreme value theory model to determine the maximal extent of the overshooting layer. However, although the convective shells in Cepheids are thin, overshooting plumes impact the full extent of the radiative layer above the convective shell and below the outer convective envelope. The crudest picture of these stars is thus one where there is a large outer convective envelope which is fully mixed by convection; this picture, however, conceals interesting and relevant details about how much the convection enhances mixing at a given radius in these stars.  The EVT statistical analysis does not provide sufficient details to allow an unprejudiced comparison between the upper and the lower overshooting layers for the convective shell.

This has motivated us to propose an alternative statistical analysis in which we separate weakly and strongly overshooting plumes and produced PDFs for each group. We then modeled the PDF of all overshooting plumes using a mixture of Gamma distributions. This allowed us to filter the contribution of the weak overshooting layer from that of the strong overshooting layer. We find that the intermittent, strong overshooting plumes correspond to $5$ to $15\%$ of all plumes and we characterized the strong overshooting distribution by its mean value and standard deviation. We defined a second characteristic length scale as the $90^{\mathrm{th}}$ percentile of the SO distribution; this length scale is comparable with the length scale obtained with the EVT, reinforcing the idea that the upper overshooting layer fills the entire radiative zone. We demonstrated that for the present simulations, the overshooting lengths thus obtained are consistent with the EVT analysis below the convection zone. This approach to measuring the extent of convective boundary mixing is new and further investigation would be required to demonstrate its utility for describing convective overshooting in simulations that include additional physical effects like rotation, magnetic fields, and pulsations.

We have used the mixture model to determine a ratio of the overshooting lengths for the overshooting layers on each side of the convective shell. For our thin and shallow convective shells, in which one length scale of convective structures dominates, the difference of overshooting above and below the shell is less than what has previously been reported in the literature. As we examine convective shells that are thicker, and deeper inside the star, the ratio of lower overshooting length to upper overshooting length drops. This ratio is always less than one for the simulations we study;  however the relatively large standard deviation associated with the mixture model allows for the possibility that at some particular instant, overshooting could be larger in the lower overshooting layer than in the upper overshooting layer.

Based on the mixture model, we have proposed a diffusion coefficient for overshooting from a convective shell, as well as a form of a diffusion coefficient for the full star that provides a more detailed picture of the enhanced mixing mechanisms in the RZ between two convective layers. Although further investigation would be necessary to fully characterize the complex mixing in overlapping overshooting layers, both the EVT and our model points out the possibility of ``super-mixing layer'' in which more efficient mixing could result in the merging of two adjacent convective zones.

The stellar structures explored in the hydrodynamic simulations in this work were not produced to systematically explore the parameter space of mass or luminosity. Furthermore, we find no clear differences other than those related to the depth and width of the convective shell between the $7\Msun$ (a) and (b) Cepheids, which are located respectively in the second and third crossing of the instability strip. In future work, a more systematic study would be required to assess a trend in the decrease of the ratio of overshooting with the depth and width of the convection zone. The present simulations are 2D, and 3D simulations would be desirable, as long as those simulations would have a radial resolution equivalent to those studied here.  A 3D simulation of shell convection might allow for the study of  instabilities of convection rolls \citep[e.g.][]{busse1979instabilities,clever1989three} in the regime relevant to stellar interiors; such instabilities could contribute to a more complete picture of convective mixing.  A better understanding of the nonlinear interactions between convection and radial pulsations in Cepheid variable stars remains necessary to deepen our theoretical knowledge of these physically complex stars; 3D simulations of the whole star, as well as simulations that include rotation and magnetic fields are necessary to expand on these results. Those directions are ones we intend to pursue.

\begin{acknowledgements}
The research leading to these results is partly supported by the ERC grants 320478-TOFU and 787361-COBOM and by the STFC Consolidated Grant  ST/Y002164/1.
This work used the DiRAC Complexity system, operated by the University of Leicester IT Services, which forms part of the STFC DiRAC HPC Facility (www.dirac.ac.uk). This equipment is funded by BIS National E-Infrastructure capital grant ST/K000373/1 and STFC DiRAC Operations grant ST/K0003259/1. DiRAC is part of the National E-Infrastructure.  This work also used the University of Exeter local supercomputer ISCA.  Computing support for this work came from the Lawrence Livermore National Laboratory (LLNL) Institutional Computing Grand Challenge program.  Part of this work was performed under the auspices of the U.S. Department of Energy by Lawrence Livermore National Laboratory under Contract DE-AC52-07NA27344.

LLNL-JRNL-2003963
\end{acknowledgements}
\bibliographystyle{aa}
\bibpunct{(}{)}{;}{a}{}{,}
\bibliography{cepheids}

\begin{appendix}

\onecolumn

\section{The structure of classical Cepheid variable stars} \label{sec:mesa}

\begin{figure*}[th!]
\begin{center}
\includegraphics[width=\textwidth]{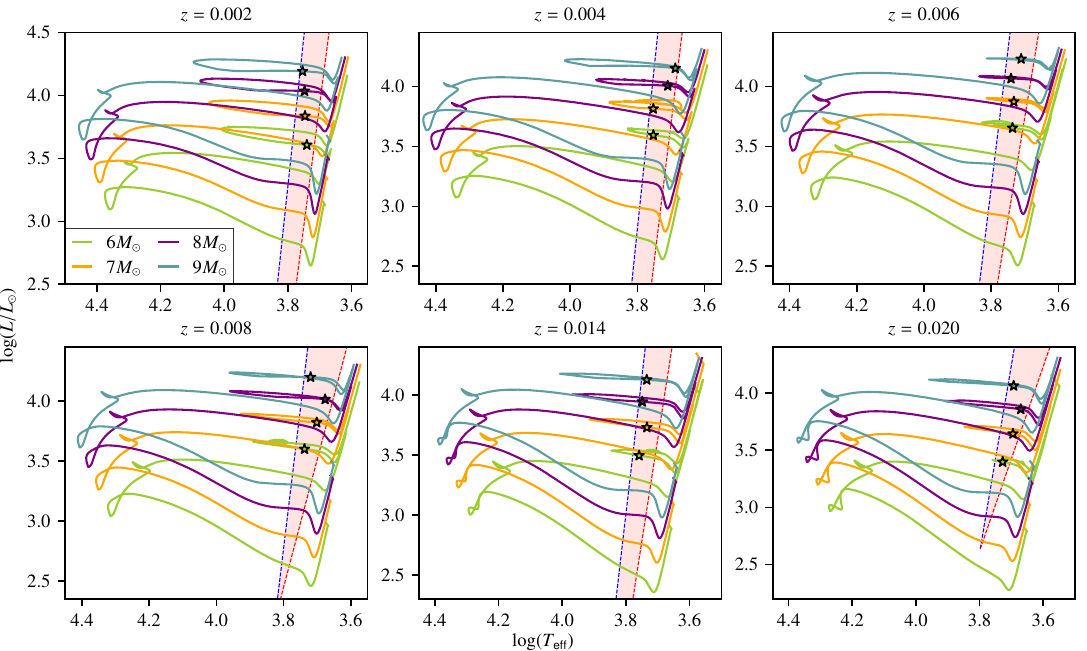}
\caption{Hertzprung--Russell (H-R) diagrams of $6$, $7$, $8$ and $9\Msun$ stars of six different metallicities. The symbols indicate structures of individual Cepheid variable stars that lie in the instability strip (pink shaded area) and that we examine further in Appendix \ref{sec:mesa}. The blue and red edges of the instability strips were estimated from the work of \cite{anderson2016effect} for $z=0.002$, $z=0.006$ and $z=0.014$, from the work of \cite{alibert1999period} for $z=0.004$, and from the theoretical relationships of \cite{bono2000classical} for $z=0.008$ and $z=0.020$. $L$ is the luminosity of the star, expressed in units of the luminosity of the sun $\Lsun$ (we use $\Lsun=3.828\times10^{33} \mathsf{\;erg.s^{-1}}$), and $T_{\mathsf{eff}}$ denotes its effective temperature.  Evolutionary tracks with different masses are distinguished by color.}
\label{HR_diagram_metallicity}
\end{center}
\end{figure*}

We use the MESA code (version 24.03.1) to understand typical structures for the interior of classical Cepheids. Our study is not exhaustive, and we limit the range of free parameters in our MESA models. We generate a grid of $48$ MESA models with $6$ different metallicities, $4$ different masses and $2$ different overshooting parameters. This study is targeted toward establishing common structures for Cepheids with parameters documented in the literature.

Classical Cepheids are intermediate mass stars that have evolved beyond the main sequence and are grouped together in the so-called instability strip of the Hertzprung--Russell (HR) diagram (indicated by the shaded areas in Fig.~\ref{HR_diagram_metallicity}) \citep{bono2005classical, anderson2014effect}. The evolutionary tracks of classical Cepheids first rapidly cross this instability strip while the hydrogen shell is still burning. When they begin to burn helium in their core, they typically execute a blue loop, crossing the instability strip a second and third time. 
Stellar structure and evolution models of intermediate mass stars that may evolve to Cepheids are sensitive to both the input parameters (including overshooting, convection, metallicity, and nuclear reaction rates) and the numerical scheme \citep[see e.g. the recent study of][]{ziolkowska2024toward}.  The occurrence and duration of blue loops produced by stellar evolution models can be impacted by the choice of input parameters.  However the duration of the blue loop tends to be smaller for lower mass stars  \citep[see, e.g.][]{walmswell2015blue}. Blue loops for stars $M<4$ solar masses (denoted $\Msun$ thereafter) that are produced by current versions of the MESA code do not cross the instability strip \citep{anderson2016effect, de2022updated, espinoza2024empirical}. For our MESA study, we therefore focus on stars in the mass range from $6$ to $9\Msun$.

\begin{figure*}[t!]
\begin{center}
\includegraphics[width=\textwidth]{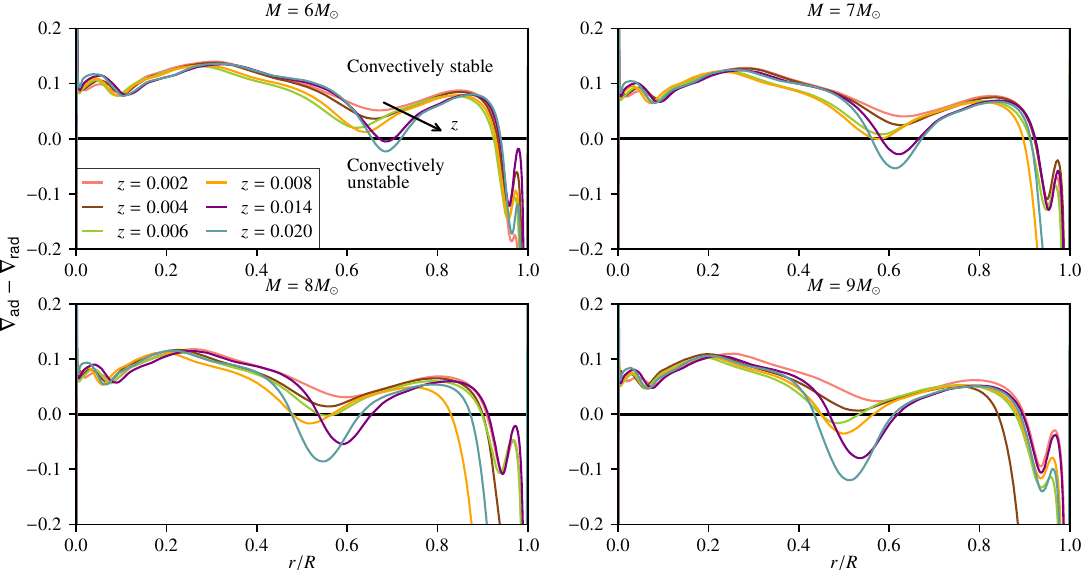}
\caption{Schwarschild discriminant as a function of stellar radius for the $6$, $7$, $8$ and $9\Msun$ stars presented in Fig.~\ref{HR_diagram_metallicity}. The radius $r$ is given in units of the total radius of each star, $R$. The arrow points in the direction of increasing metallicity.  Profiles with different metallicities are distinguished by color.  \label{schwarschild}}
\end{center}
\end{figure*}

Based on inlists published by \cite{espinoza2022period}, we produce evolutionary tracks in this mass range for a range of stellar metallicities between $z=0.002$ and $z=0.020$. This range of metallicities has been documented in various Cepheid populations \citep[see e.g., ][]{baraffe1998cepheid, anderson2016effect, chiosi1992instability, alibert1999period, hocde2024pulsation}.  For each mass and metallicity, the helium mass fraction $y$ is set equal to $0.256$ and we use a mixing length parameter of $\alpha_{\mathsf{mlt}}=1.88$. Exponential core and envelope overshooting is used at the convective boundaries. For half of the evolutionary tracks produced, we set the overshooting parameter to $f_{\mathsf{ov}} = 0.009$; for the other half we set  $f_{\mathsf{ov}} = 0.019$ so that we can examine the effect of overshooting in a binary way. For these parameters, for each mass and metallicity, we find at least one evolutionary track that executes a blue loop large enough to cross the instability strip twice. The evolutionary tracks of $24$ of these stars are plotted in Fig.~\ref{HR_diagram_metallicity} with symbols indicating selected models of Cepheids that we examine in detail below. The blue and red edges of the instability strips were estimated from the work of \cite{anderson2016effect} for $z=0.002$, $z=0.006$ and $z=0.014$, from the work of \cite{alibert1999period} for $z=0.004$, and from the theoretical relationships of \cite{bono2000classical} for $z=0.008$ and $z=0.020$.

\begin{figure*}[h!]
\begin{center}
\includegraphics[width=\textwidth]{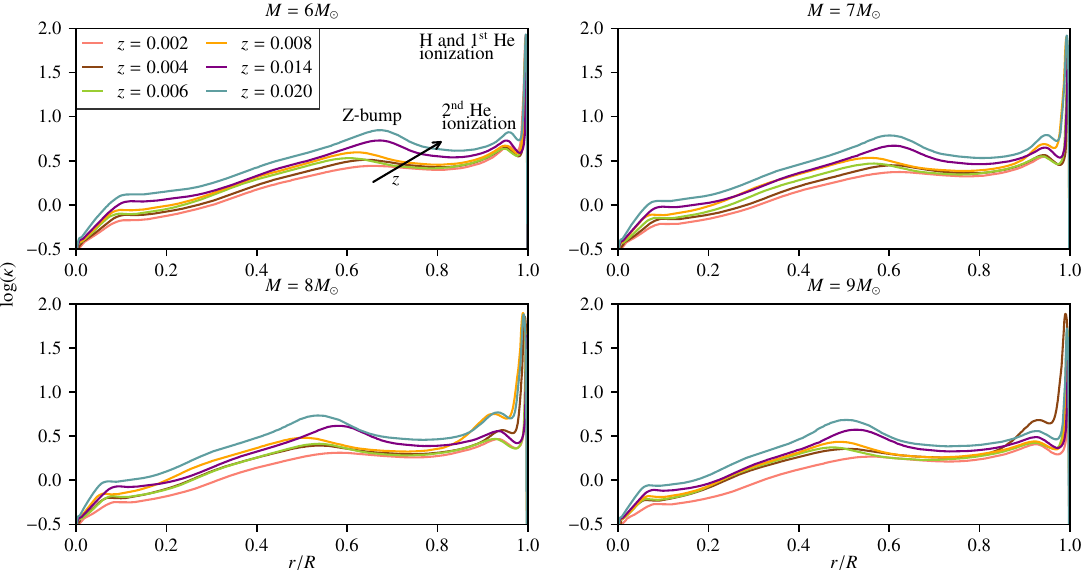}
\caption{Radial profiles of opacity ($\kappa$) for the $6$, $7$, $8$ and $9\Msun$ stars presented in Fig.~\ref{HR_diagram_metallicity}. The arrow points in the direction of increasing metallicity.  Profiles with different metallicities are distinguished by color.  \label{opacity}}
\end{center}
\end{figure*}

\twocolumn

\begin{figure}[t!]
\begin{center}
\includegraphics[width=\columnwidth]{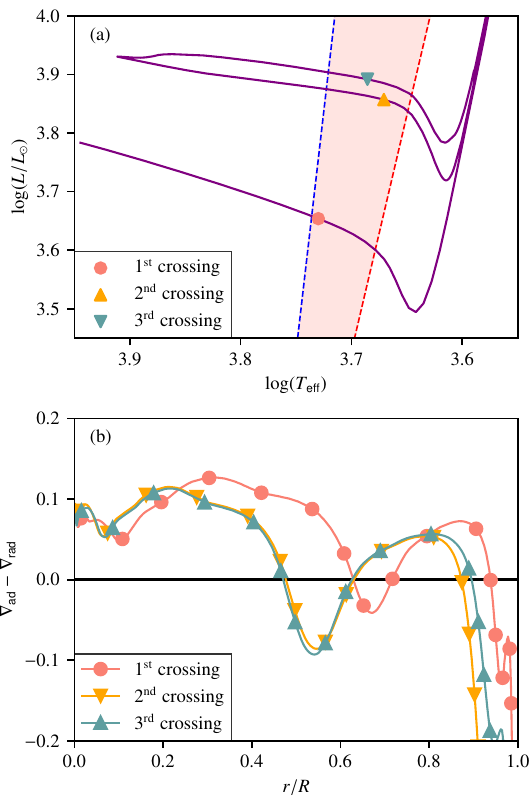}
\caption{Cepheids for an $8\Msun$ star that has metallicity $z=0.020$ (also shown in Fig.~\ref{HR_diagram_metallicity}): (a) the H-R diagram showing that this star crosses the instability strip three times. Symbols denote three stellar models corresponding to the Schwarzschild discriminants in (b). \label{comp_crossings}}
\end{center}
\end{figure}

\begin{figure}[h!]
\begin{center}
\includegraphics[width=\columnwidth]{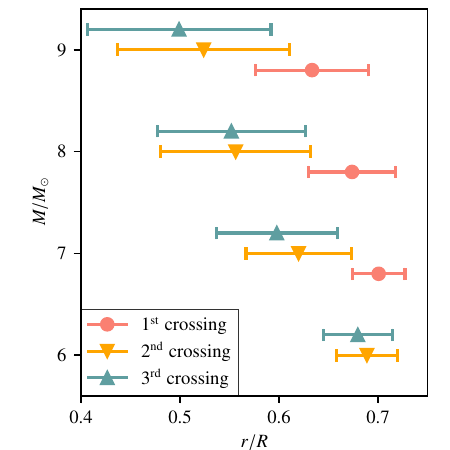}
\caption{Location and width of the convective shell for each of the three crossings of the instability, for the $6$, $7$, $8$ and $9\Msun$ stars at $z=0.020$.}
\label{comp_crossings_conv_shell}
\end{center}
\end{figure}

Because our interest is stellar convection, we focus on the presence of convectively unstable zones in the interior of Cepheids, and their dependence on mass and metallicity. To characterize whether a layer of the stellar interior is stable against convection, we use the Schwarzschild criterion \citep{schwarzschild1958structure}.  For convectively unstable layers, the Schwarzschild discriminant, defined as the difference between the adiabatic and the radiative temperature gradients, $\DPS \nabla_{\mathsf{ad}} - \nabla_{\mathsf{rad}}$ is negative.
We first examine several stars that are crossing the instability strip for the second time.
On the evolutionary tracks in Fig.~\ref{HR_diagram_metallicity}, several points are indicated; for the stellar structures at these points, we plot the Schwarzschild discriminant against the star's internal radius (see Fig.~ \ref{schwarschild}). For each mass and metallicity, these Cepheids exhibit a convective core that is so small that the curve visibly merges with the left axis, and a small outer convective envelope just below the photosphere. For the $8\Msun$ Cepheid with $z=0.020$, we find that the convective boundary of the convective core is located at $0.65\%$ of the stellar radius, but represents $9\%$ of the stellar mass. In contrast, the outer convection zone corresponds to the outer $5 \%$ of the star's radius.

The Schwarzschild discriminant also exhibits a dip in the interior of the star, in the range $0.4<r/R<0.7$ (see Fig.~\ref{schwarschild}).   For sufficiently high metallicity, this dip corresponds to the appearance of a convectively unstable layer.   The position of this dip corresponds to the so-called Z-bump \citep{iglesias1992spin, rogers1992rosseland, seaton1993fitting, seaton1994opacities, daszynska2019testing} in the opacity profiles (see Fig.~\ref{opacity}). The Z-bump is a local maximum in the Rosseland mean opacity that occurs around a temperature of $2\times 10^5$K; this phenomenon is caused by a large number of bound-bound but also bound-free transitions in iron-group elements.  The metallicity of the star influences the width and depth of this convection zone; as the metallicity decreases, this convective shell becomes thinner until it completely disappears at the lowest metallicities that we consider (see Fig.~\ref{schwarschild}). Based on the grid of MESA models that we produce, we find that this convective shell is wider and deeper for more massive stars.  Above a minimum mass, the convective shell also exists for lower metallicities. Thus, for the $6$ and $7\Msun$ stars, this convectively unstable region exists close to solar abundance and vanishes at lower $z$; the Schwarzschild discriminant is positive and the opacity Z-bump is less pronounced in these cases. The mass of the star affects the location of the convective shell, although the metallicity appears to have only a limited influence on it. As the mass of the Cepheid increases, the convective shell is shifted deeper in the star. Although further investigation would be required to fully characterize this inner convective shell, the combination of a tiny convective core, an inner convective shell, and an outer convective envelope is a common structure for Cepheids on their second crossing of the instability strip, provided that they have sufficiently high mass and metallicity. 

For the first and third crossings of the instability strip (see Fig.~\ref{comp_crossings}) we find similar results.   During all three crossings, a convective shell appears in the Z-bump region, (see Fig.~\ref{comp_crossings} for a comparison of structures for the three crossings of a $8\Msun$ star).  For stars that are on the first and third crossings of the instability strip, the same dependence on mass and metallicity exists; this convective shell becomes thinner as the mass decreases, and disappears at low mass and low metallicity (see Fig.~\ref{comp_crossings_conv_shell}). This comparison of the width and location of the convective shell for the $6$, $7$, $8$ and $9\Msun$ stars at $z=0.020$, for each of the three crossings of the instability strip, exhibits a clear trend.  No convective shell exists for the $6\Msun$ star during the first crossing, even at the highest metallicity studied, but a convective shell appears during this star's second and third crossings of the instability strip. The location of the unstable region slightly changes as one moves along the evolutionary track in the instability strip, and it is shallower and smaller at the first crossing compared with the second and third crossings. As a consequence, the convective shell tends to disappear at higher metallicities than for the last two crossings. As illustrated in Figs.~\ref{comp_crossings} and \ref{comp_crossings_conv_shell}, few differences are observed between the second and third crossings.  As we consider stars of higher mass, the convective shell is both wider and deeper, for all three crossings of the instability strip. 

\end{appendix}

\end{document}